\documentclass{aastex6}
\usepackage{graphicx}

\fullcollaborationName{The Friends of AASTeX Collaboration}

\begin{document}


\title{Possible Counter Rotation between the Disk and Protostellar Envelope\\
around the Class I Protostar IRAS 04169+2702}



\author{Shigehisa Takakuwa\altaffilmark{1,2}, Yusuke Tsukamoto\altaffilmark{1},
Kazuya Saigo\altaffilmark{3}, \& Masao Saito\altaffilmark{4}}
\altaffiltext{1}{Department of Physics and Astronomy, Graduate School of Science and Engineering,
Kagoshima University, 1-21-35 Korimoto, Kagoshima, Kagoshima 890-0065, Japan;
takakuwa@sci.kagoshima-u.ac.jp}
\altaffiltext{2}{Academia Sinica Institute of Astronomy and Astrophysics,
P.O. Box 23-141, Taipei 10617, Taiwan}
\altaffiltext{3}{ALMA Project Office, National Astronomical Observatory of Japan, Osawa 2-21-1,
Mitaka, Tokyo 181-8588, Japan}
\altaffiltext{4}{TMT-J Project Office, National Astronomical Observatory of Japan, Osawa 2-21-1,
Mitaka, Tokyo 181-8588, Japan}


\begin{abstract}
We present results from our SMA observations and data analyses of the SMA archival data
of the Class I protostar IRAS 04169+2702.
The high-resolution ($\sim$0$\farcs$5) $^{13}$CO (3--2) image cube shows
a compact ($r \lesssim 100$ au) structure with a northwest (blue)
to southeast (red) velocity gradient, centered on the 0.9-mm dust-continuum emission.
The direction of the velocity gradient is orthogonal to the axis of the molecular outflow
as seen in the SMA $^{12}$CO (2--1) data. A similar gas component is seen
in the SO (6$_5$--5$_4$) line.
On the other hand, the C$^{18}$O (2--1) emission traces a more extended ($r \sim$400 au)
component with the opposite, northwest (red) to southeast (blue) velocity
gradient. Such opposite velocity gradients in the different molecular lines
are also confirmed from direct fitting to the visibility data.
We have constructed models of a forward-rotating and counter-rotating Keplerian disk
and a protostellar envelope, including the SMA imaging simulations.
The counter-rotating model could better reproduce the observed velocity channel maps,
although we could not obtain statistically significant fitting results.
The derived model parameters are; Keplerian radius of 200 au,
central stellar mass of 0.1 $M_{\odot}$,
and envelope rotational and infalling velocities of 0.20 km s$^{-1}$
and 0.16 km s$^{-1}$, respectively.
One possible interpretation for these results
is the effect of the magnetic field in the process of disk formation around
protostars, $i.e.$, Hall effect.
\end{abstract}




\section{Introduction} \label{sec:intro}

Recent observational efforts have been finding Keplerian disks around
not only T-Tauri stars ($e.g.$, Simon et al. 2000; Williams \& Cieza 2011)
but also protostars \cite{tob12,har14,lin14,aso15}.
The disks around protostars are considered to be under the formation
and growth process of “protoplanetary disks”, precursors of planetary systems.
Studies of disk formation processes around protostars
are thus important to understand the
initial condition of planet formation.

Theoretical studies have suggested 
that magnetic fields play a vital role 
in disk formation from protostellar envelopes \cite{li14,mac16}.
The magnetic field connects the inner regions and the outer envelopes
and efficiently transfers the angular momentum from the inner 
to outer regions. This process is known as magnetic 
braking \cite{gil74,mou85,all03}.
If an ionization degree of the cloud cores is high enough and
ideal magnetohydrodynamics (MHD) approximation is valid,
the magnetic braking is so efficient and almost completely suppresses
the circumstellar disk formation in cloud cores
under a typical magnetic-field strength \cite{mel08,hen09}.
The ionization degree of the real cloud cores is, however,
very low \cite{ume90,nis91,cas98,nak02},
and non-ideal MHD effects (Ohmic diffusion, Hall effect, 
and ambipolar diffusion) caused by low conductivity of the gas
must play a role during the cloud core collapse.
It has been suggested that the Ohmic and ambipolar diffusion decouples
the magnetic field and the gas at $\rho>10^{-12}~{\rm g~cm^{-3}}$ and 
circumstellar disk formation is enabled even at the
very early phase of star formation \cite{ma11a,tsu15a,tom15,mas16}.
Recent theoretical simulations show that the Hall effect imprints
the characteristic velocity structure in the envelope and disk, $i.e.$,
a flip of the rotation velocity, or counter rotation
\cite{kra11,li11,tsu15b,wur16,tsu17}.

Another important physical mechanism that controls disk formation
is turbulence. Theoretical simulations show that turbulence can reduce
the magnetic braking through magnetic diffusions and reconnections,
and add additional angular momenta \cite{san12,joo13,sei13,mat17}.
Such turbulent effects promote disk formation around protostars.
Furthermore, the rotational axis of the formed disk can be misaligned
from that of the surrounding protostellar envelope \cite{mat17}.

It has been difficult, however, to observationally identify
gas motions which are indeed controlled by magnetic fields or turbulence in
disk-forming regions. Detailed observational comparisons
of gas motions from dense cores, envelopes, to central disks
are essential to tackle this problem \cite{tob11,har14}.
In particular,
radial rotational profiles from the envelope to the inner disk
have been measured observationally \cite{yen13,har14,har15,aso15}.
Those observations show that the radial rotational profiles in the
envelopes and disks can be approximated to be $v_{rot} \sim r^{-1}$
and $\sim r^{-0.5}$, that is, rotation
with the conserved specific angular momenta and Keplerian rotation,
respectively.
While these results apparently show that the magnetic or turbulent effect
on the gas motions is not significant, the rotational profiles
and the power-law indices derived from these observations are
not accurate enough to be directly compared with those from
theoretical simulations.
%

A more straightforward observational signature of
such effects is desirable. We consider that change of the rotational
axes between the central disk and the outer envelope, or even the flip
of the rotational vectors, is an intriguing tracer for the effect
of magnetic field and turbulence.
In this paper, we report
SubMillimeter Array (SMA)\footnote{The SMA is a joint project between the Smithsonian Astrophysical Observatory
and the Academia Sinica Institute of Astronomy and Astrophysics and is funded by the
Smithsonian Institution and the Academia Sinica.}
observations of IRAS 04169+2702 at 330 GHz as well as data analyses of the SMA archival data at 230 GHz.
IRAS 04169+2702 (hereafter I04169) is a Class I protostar
($L_{bol}$ = 0.76 $L_{\odot}$; $T_{bol}$ = 133 K)
\cite{ken93a,ken93b,you03} located in the molecular filament of the B213 / L1495 region
at $d$ = 140 pc \cite{hac13,taf15}. The protostar is associated with
a $\sim$20000 AU scale, $\sim$1 $M_{\odot}$ dense core as seen in the
1.3-mm and 850 $\micron$ dust-continuum \cite{mot01,you03} and N$_{2}$H$^{+}$ (1--0) emission \cite{tat04}.
%
%
%
%
Previous millimeter interferometric observations of I04169
in the C$^{18}$O (1--0) \cite{oha97} and H$^{13}$CO$^{+}$ (1--0) lines
\cite{sai01} have found a $r \sim$1000 au scale
protostellar envelope elongated along the northwest to southeast direction
(P.A. = 154$\degr$) with an inclination angle of $i \sim60\degr$.
In the C$^{18}$O (1--0) line, the southeastern part of the envelope
is blueshifted ($\sim$-0.8 km s$^{-1}$ from $v_{sys}$ = 6.8 km s$^{-1}$)
and northwestern part redshifted ($\sim$+0.8 km s$^{-1}$),
and this velocity gradient along the major
axis is regarded as the rotation of the envelope \cite{oha97}.
CARMA 1.3-mm continuum and Keck I-band imaging of I04169 exhibit
a small, almost unresolved ($\lesssim 1\arcsec$) dusty disk without
any scattered light \cite{eis12}.

From the SMA data of I04169, we have found possible observational evidence
for a counter rotation between the protostellar envelope and circumstellar disk,
which will be shown in the rest of the present paper. In Section 2, we shall describe our
new SMA observations and archival data, and calibrations and imaging of those data.
In Section 3, the 0.9-mm and 1.3-mm continuum, $^{12}$CO (2--1), $^{13}$CO (2--1; 3--2),
C$^{18}$O (2--1), and SO (6$_5$--5$_4$) results are presented and compared.
Section 4 describes our modeling efforts to reproduce the observed velocity structures
with the SMA. In Section 5.1 we discuss physical origin
of the observed gas motions around I04169 and in section 5.2 implications of these results.

\section{SMA Observations and Data Reduction} \label{sec:obs}

SMA Observations of I04169 at 330 GHz were made on 2014 December 19
and 2015 February 6 with its very extended and extended configurations, respectively.
Details of the SMA are described by Ho et al. (2004).
The correlator covered the
$^{13}$CO ($J$=3--2; 330.588 GHz), C$^{18}$O ($J$=3--2; 329.331 GHz)
and the CS ($J$=7--6; 342.883 GHz) lines,
and the spectral windows (``chunks") of the correlator
with 512 channels were assigned to these lines. The bandwidth of one chunk is 82 MHz,
and thus the spectral resolution of these lines is 203.125 kHz.
In each sideband there were a total of 48 chunks, and
all the chunks at both sidebands except for those assigned to the lines
were combined to make a single continuum channel.
The central frequency of the continuum channel is $\sim$336.933 GHz ($\lambda \sim$0.89 mm),
and hereafter the continuum emission is called as the 0.9-mm continuum emission.
The minimum projected baseline length was $\sim$29 $k\lambda$ at the $^{13}$CO (3--2) frequency,
and for a Gaussian emission distribution with an FWHM of $\sim$5$\farcs$7 ($\sim$800 AU),
the peak flux density recovered is $\sim$10$\%$ of the peak flux density of the Gaussian \cite{wil94}.
The uncertainty in the absolute flux calibration is inferred to be $\sim$30$\%$.
Table \ref{330obs} summarizes the observational parameters.

SMA archival data of I04169 at 230 GHz
taken in 2011 December 15 and 2012 February 15 with the compact and extended
configurations, respectively, were retrieved and re-calibrated.
The data include the $^{12}$CO ($J$=2--1),
$^{13}$CO ($J$=2--1), C$^{18}$O ($J$=2--1), and the SO ($J_N$=6$_5$--5$_4$) lines.
The SMA chunks with 512 channels were allocated to the three CO isotopic lines,
while a chunk with 128 channels to the SO line.
The central frequency of the continuum channel is $\sim$225.506 GHz ($\lambda \sim$1.33 mm),
and hereafter the continuum emission is called as the 1.3-mm continuum emission.
Unfortunately, the spectral setting
was not optimized to the C$^{18}$O (2--1) line, and the C$^{18}$O spectral region
with $V_{LSR} >$ 7.6 km s$^{-1}$ was dropped out of the chunk with 512 channels.
In contrast to the 330 GHz observations described above, the 230 GHz observations
cover a factor $\sim$3 shorter spacings, and the minimum projected baseline length
($\sim$9.9 $k\lambda$)
at the C$^{18}$O (2--1) frequency implies that for a Gaussian emission distribution
with an FWHM of $\sim$16$\farcs$7 ($\sim$ 2300 AU),
the peak flux density recovered is $\sim$10$\%$ of the peak flux density
of the Gaussian \cite{wil94}.
On the other hand, the angular resolution of the 230 GHz observations is limited to
$\sim$1$\farcs$5, while that of the 330 GHz observations is as high as $\lesssim$0$\farcs$5.
These facts imply that the 230 GHz observations are more suitable to investigate the
protostellar envelope around I04169, while the 330 GHz observations to study the central
disk. Table \ref{230obs} summarizes the parameters of the 230 GHz observations.

The raw visibility data were calibrated with an IDL-based reduction package, MIR \cite{sco93},
and the calibrated visibility data were Fourier-transformed and CLEANed with MIRIAD \cite{sau95}.
Depending on the intensities of the observed molecular lines different visibility weightings
were adopted to construct the images. The Natural weighting and 0$\farcs$3 tapering were applied
to the $^{13}$CO (3--2) visibility data, to recover enough fluxes and to construct the
image cube with a sufficient signal-to-noise ratio. With the same visibility weighting and tapering
the C$^{18}$O (3--2) emission is only marginally detected above 4$\sigma$ (1$\sigma$ = 0.13 J beam$^{-1}$)
at $V_{\rm LSR}$ = 5.8 and 6.0 km s$^{-1}$, and there is no CS (7--6) emission detected.
The Natural weighting and 2$\farcs$0 tapering are applied to the C$^{18}$O (2--1) visibility data
to recover enough fluxes of the protostellar envelope.
The adopted visibility weightings and the resultant beam sizes and the noise levels
of the different molecular-line data are summarized in Table \ref{lineobs}.
Regarding the continuum data both the Natural and Uniform weightings were adopted
to investigate the changes of the continuum images with the two extreme weightings
(see the next section).

\begin{deluxetable}{lcc}
\tabletypesize{\scriptsize}
\tablecaption{Parameters for the SMA Observations of I04169 at 330 GHz \label{330obs}}
\tablewidth{0pt}
\tablehead{\colhead{Parameter} & \multicolumn{2}{c}{Value}\\
\cline{2-3}
\colhead{} & \colhead{2014 December 19} & \colhead{2015 February 6} }
\startdata
Number of Antennas &7              &5 \\
Configuration      &Very Extended  &Extended \\
Right ascension (J2000.0)
   & \multicolumn{2}{c}{04$^{\rm h}$ 19$^{\rm m}$ 58$^{\rm s}$.45}\\
Declination (J2000.0)
   & \multicolumn{2}{c}{27$^{\circ}$ 09$\arcmin$ 57\farcs1}\\
Primary Beam HPBW& \multicolumn{2}{c}{$\sim$37$\arcsec$}\\
Baseline Coverage & \multicolumn{2}{c}{29 - 561 (k$\lambda$)}\\
Conversion Factor ($^{13}$CO) & \multicolumn{2}{c}{1 (Jy beam$^{-1}$) = 38.7 (K)}\\
Continuum Bandwidth &  \multicolumn{2}{c}{7.87 GHz}\\
Flux Calibrator &3c84 &Callisto\\
Gain Calibrator & \multicolumn{2}{c}{3c84, 3c111}\\
Flux (3c84)     &7.6 Jy   &7.4 Jy \\
Flux (3c111)    &0.93 Jy  &0.91 Jy \\
Passband Calibrator    &\multicolumn{2}{c}{3c279}  \\
System Temperature &$\sim$400 - 1200 K &$\sim$300 - 700 K \\
\enddata
\end{deluxetable}

\begin{deluxetable}{lcc}
\tabletypesize{\scriptsize}
\tablecaption{Parameters for the SMA Archival Data of I04169 at 230 GHz \label{230obs}}
\tablewidth{0pt}
\tablehead{\colhead{Parameter} & \multicolumn{2}{c}{Value}\\
\cline{2-3}
\colhead{} & \colhead{2011 December 15} & \colhead{2012 February 15} }
\startdata
Number of Antennas &\multicolumn{2}{c}{8} \\
Configuration      &Compact  &Extended \\
Right ascension (J2000.0)
   & \multicolumn{2}{c}{04$^{\rm h}$ 19$^{\rm m}$ 58$^{\rm s}$.44}\\
Declination (J2000.0)
   & \multicolumn{2}{c}{27$^{\circ}$ 09$\arcmin$ 57\farcs0}\\
Primary Beam HPBW& \multicolumn{2}{c}{$\sim$56$\arcsec$}\\
Baseline Coverage & \multicolumn{2}{c}{10 - 164 (k$\lambda$)}\\
Conversion Factor (C$^{18}$O) & \multicolumn{2}{c}{1 (Jy beam$^{-1}$) = 2.83 (K)}\\
Continuum Bandwidth &  \multicolumn{2}{c}{6.48 GHz}\\
Flux Calibrator &Callisto   &Uranus\\
Gain Calibrator & \multicolumn{2}{c}{3c84, 3c111}\\
Flux (3c84)     &8.8 Jy   &8.4 Jy \\
Flux (3c111)    &2.3 Jy  &2.4 Jy \\
Passband Calibrator &3c279 &3c279, Mars \\
System Temperature &   $\sim$200 - 500 K          & $\sim$150 - 300 K    \\
\enddata
\end{deluxetable}

\begin{deluxetable}{lcclcc}
\tabletypesize{\scriptsize}
\tablecaption{Summary of the Line Images\label{lineobs}}
\tablewidth{0pt}
\tablehead{\colhead{Line} &\colhead{Frequency} &\colhead{Weighting}
&\colhead{Beam (P.A.)} &\colhead{Velocity Resolution} &\colhead{rms} \\
\colhead{} &\colhead{(GHz)} &\colhead{} &\colhead{}
&\colhead{(km s$^{-1}$)} &\colhead{(Jy beam$^{-1}$)}}
\startdata
$^{13}$CO ($J$=3--2)    &330.587965 &Natural, 0$\farcs$3 taper &0$\farcs$59$\times$0$\farcs$49 (86$\degr$) &0.18 &0.12 \\
C$^{18}$O ($J$=2--1)    &219.560358 &Natural, 2$\farcs$0 taper &3$\farcs$09$\times$2$\farcs$90 (-69$\degr$) &0.28 &0.099 \\
$^{12}$CO ($J$=2--1)    &230.538000 &Uniform                   &1$\farcs$54$\times$1$\farcs$37 (-80$\degr$) &0.26 &0.14 \\
$^{13}$CO ($J$=2--1)    &220.398684 &Natural                   &1$\farcs$82$\times$1$\farcs$52 (-88$\degr$) &0.28 &0.060 \\
SO ($J_N$=6$_5$--5$_4$) &219.949433 &Natural                   &1$\farcs$82$\times$1$\farcs$52 (-88$\degr$) &1.11 &0.028 \\
\enddata
\end{deluxetable}

\section{Results} \label{sec:res}
\subsection{Continuum and CO Outflow}

\begin{figure}[ht!]
\figurenum{1}
\epsscale{1}
\begin{center}
\includegraphics[scale=0.5,angle=0]{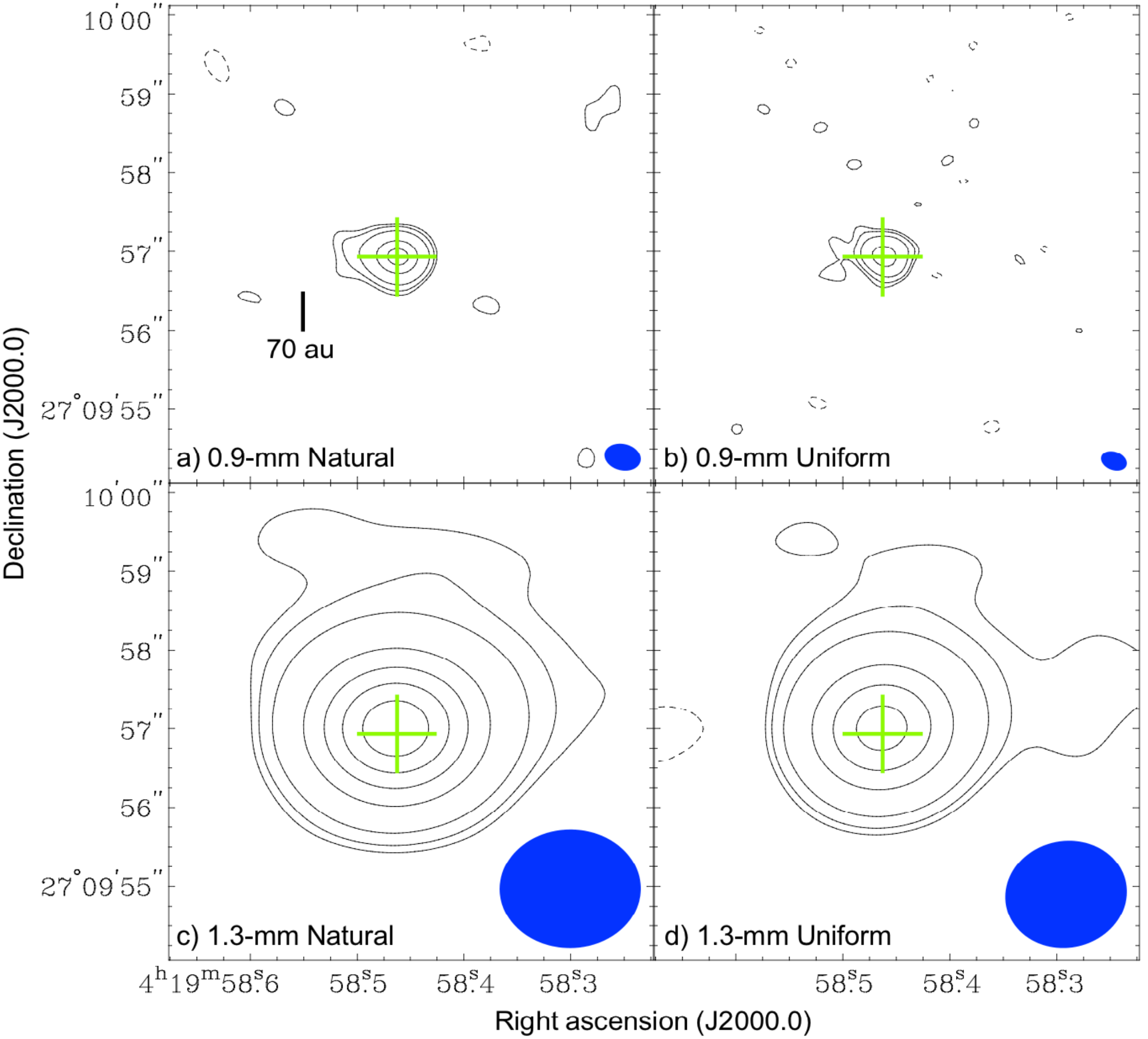}
\end{center}
\caption{SMA continuum images of I04169. Contour levels are
3$\sigma$, 5$\sigma$, 10$\sigma$, 30$\sigma$, 50$\sigma$, 70$\sigma$, 90$\sigma$,
110$\sigma$, where 1$\sigma$ levels are summarized
in Table \ref{contobs}.
Crosses show the centroid position of the 0.9-mm continuum image
with the Natural weighting, which we regard as the position of the protostar.
Filled ellipses at the bottom-right corners show the synthesized beams,
which are also summarized in Table \ref{contobs}.
\label{fig:contall}}
\end{figure}

Figure \ref{fig:contall} shows the 0.9-mm and 1.3-mm dust-continuum images of I04169
at the Natural and Uniform weightings, and Table \ref{contobs}
summarizes their image quantities.
The 1.3-mm continuum emission is not resolved with the present angular resolution
of $\sim$1$\farcs$5. Even with the higher-resolution observations at 0.9-mm
the dust-continuum emission is barely resolved, and the beam-deconvolved
emission sizes are as small as the beam sizes (Table \ref{contobs}).
In the Natural-weighted 0.9-mm
continuum image there appears a marginal emission extension to the east,
but this component is degraded in the Uniform-weighted image.
These results indicate that from the continuum images
it is not possible to identify the direction of the major axis of the
circumstellar disk. In the following, we regard
the centroid position of the 0.9-mm continuum image with the Natural weighting
measured from the 2-dimensional Gaussian fitting as the protostellar position
(04$^{\rm h}$ 19$^{\rm m}$ 58$\fs$463, 27$\degr$ 09$\arcmin$ 56$\farcs$936).

From the continuum flux densities derived from the Natural-weighted images at the two frequency bands,
plus the continuum flux density at 2.7 mm ($\sim$17 mJy; Ohashi et al. 1997),
the spectral index $\alpha$ is calculated to be $\alpha \sim$2.19$\pm$0.25. 
The measured $\alpha$ value indicates a small $\beta$ value, $i.e.$, $\beta \sim$0--0.5,
suggesting dust growths.
The mass of the dusty component around I04169
($\equiv M_{d}$) is estimated from the measured 0.9-mm continuum flux density ($\equiv S_{\nu}$)
at the Natural weighting as;
\begin{equation}
M_{d}=\frac{S_{\nu}d^2}{\kappa_{\nu} B_{\nu}(T_d)},
\end{equation}
where $\nu$ is the frequency, $d$ the distance, $B_{\nu}(T_d)$ the Planck function
for dust at a temperature $T_{d}$, and $\kappa_{\nu}$ the dust opacity per unit gas + dust mass
on the assumption of a gas-to-dust mass ratio of 100.
The mass opacity at 0.9 mm is calculated to be $\kappa_{0.9mm}$ = 0.053 cm$^{2}$ g$^{-1}$
from $\kappa_{\nu}$ = $\kappa_{\nu_{0}}$($\nu$/$\nu_{0}$)$^{\beta}$,
$\kappa_{250~\mu m}$=0.1 cm$^{2}$ g$^{-1}$ \cite{hil83}, and $\beta$=0.5. 
For $T_d$ = 10 - 30 K, the mass of the dusty circumstellar material is calculated to be
0.0042 - 0.024 $M_{\odot}$.
The adopted dust mass opacity
is a factor 3 higher than that of Ossenkopf \& Henning (1994) for grains
with thin ice mantles coagulated at a density of 10$^{6}$ cm$^{-3}$
($\kappa_{0.9mm}$ = 0.018 cm$^{2}$ g$^{-1}$).
Thus, adopting the dust mass opacity
by Ossenkopf \& Henning (1994) yields a factor 3 higher mass.

\begin{deluxetable}{cccccc}
\tabletypesize{\scriptsize}
\tablecaption{1.3-mm and 0.9-mm Continuum Results of I04169\label{contobs}}
\tablewidth{0pt}
\tablehead{\colhead{$\lambda$} &\colhead{Weighting} &\colhead{Beam (P.A.)}
&\colhead{rms (mJy)} &\colhead{Flux Density (Jy)\tablenotemark{a}} &\colhead{Deconvolved Size\tablenotemark{a}}}
\startdata
1.3 mm &Natural &1$\farcs$76$\times$1$\farcs$50 (-89$\degr$) &0.84 &0.088 &\nodata  \\
       &Uniform &1$\farcs$52$\times$1$\farcs$35 (-78$\degr$) &1.1  &0.085 &\nodata  \\
0.9 mm &Natural &0$\farcs$46$\times$0$\farcs$34 (76$\degr$)  &2.2  &0.189 &0$\farcs$30$\times$0$\farcs$23 (-70$\degr$) \\
       &Uniform &0$\farcs$33$\times$0$\farcs$23 (68$\degr$)  &2.2  &0.180 &0$\farcs$27$\times$0$\farcs$25 (-46$\degr$) \\
\enddata
\tablenotetext{a}{Derived from the 2-dimensional Gaussian fitting to the continuum images shown in Figure \ref{fig:contall}.}
\end{deluxetable}

\begin{figure}[ht!]
\figurenum{2}
\epsscale{1}
\begin{center}
\includegraphics[scale=0.5,angle=-90]{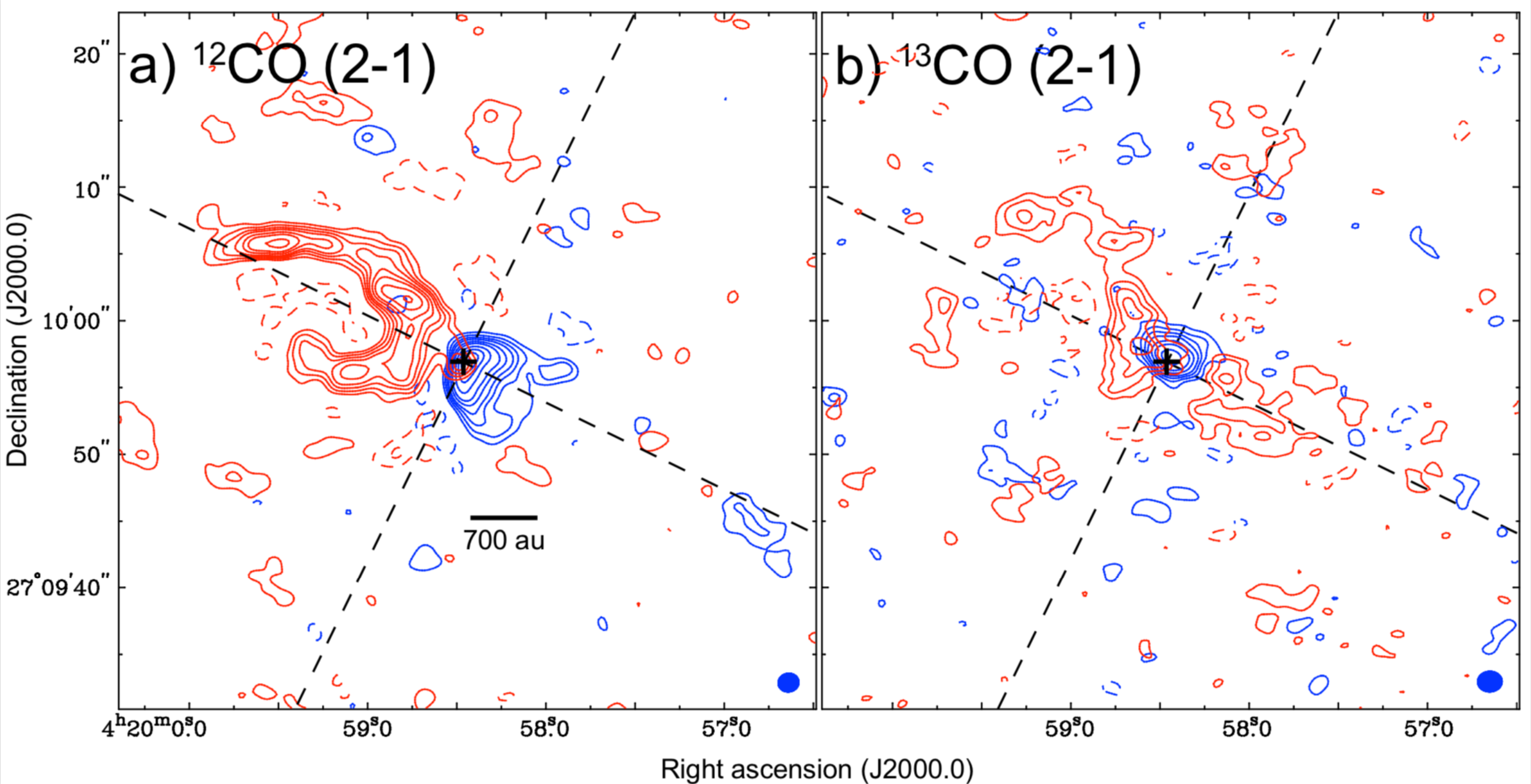}
\end{center}
\caption{a) SMA images of the blueshifted ($V_{\rm LSR}$ = 0.13 -- 5.94 km s$^{-1}$)
and redshifted (7.79 -- 15.19 km s$^{-1}$) $^{12}$CO (2-1) emission in I04169.
Contour levels are 3$\sigma$, 5$\sigma$, 7$\sigma$, 9$\sigma$, 12$\sigma$, 15$\sigma$,
and then in steps of 5$\sigma$ (1$\sigma$ = 0.17 Jy beam$^{-1}$ km s$^{-1}$ and
0.20 Jy beam$^{-1}$ km s$^{-1}$ for the blueshifted and redshifted emission.).
A cross show the position of the protostar,
and a filled ellipse at the bottom-right corner the synthesized beam
(1$\farcs$54$\times$1$\farcs$37; P.A.=-80$\degr$).
Dashed lines denote the position angles of 154$\degr$ and 64$\degr$,
which are regarded as the major and minor axes of the envelope around
I04169, respectively.
b) SMA images of the blueshifted (3.21 -- 6.80 km s$^{-1}$) and
redshifted (6.80 -- 10.39 km s$^{-1}$) $^{13}$CO (2-1)
emission in I04169. Contour levels are 3$\sigma$, 5$\sigma$, 7$\sigma$, 9$\sigma$, 12$\sigma$, 15$\sigma$,
and then in steps of 5$\sigma$ (1$\sigma$ = 0.060 Jy beam$^{-1}$ km s$^{-1}$).
A filled ellipse at the bottom-right corner shows the synthesized beam
(1$\farcs$82$\times$1$\farcs$52; P.A.=-88$\degr$).
\label{fig:outflow}}
\end{figure}

Figure \ref{fig:outflow} shows distributions of the blueshifted and redshifted
$^{12}$CO (2--1) and $^{13}$CO (2--1) emission in I04169. For comparison,
the directions of the associated outflow measured from the $^{12}$CO (1--0) observations
(P.A.=64$\degr$; dashed line along the NE-SW direction)
and the major axis of the $r\sim$1000-au scale protostellar envelope as seen
in the C$^{18}$O (1--0) emission (P.A.=154$\degr$; dashed line along the NW-SE direction)
are also shown \cite{oha97}.
These two axes are orthogonal with each other.
The redshifted $^{12}$CO (2--1) emission exhibits a tilted
$U$-shaped feature with its symmetric axis consistent with the $^{12}$CO (1--0)
outflow axis. The blueshifted $^{12}$CO (2--1) emission shows a tilted $V$-shaped
feature with its apex close to the protostellar position.
The symmetric axis of the blueshifted emission is also consistent with the
$^{12}$CO (1--0) outflow axis. These results indicate that the $^{12}$CO (2--1) emission
also traces the molecular outflow driven from I04191.
There are other possible emission components, such as
the 7$\sigma$ redshifted peak to the northeast of the protostar
and the 5$\sigma$ blueshifted peak to the southwest, as
they appear consecutively in the velocity channel maps.
The redshifted $^{13}$CO (2--1) emission to the east of the protostar appears to trace
the outflow component too. On the other hand, the blueshifted $^{13}$CO (2--1) emission
is compact and located to the northwest of the protostar, and thus its origin is likely
distinct from that of the blueshifted $^{12}$CO (2--1) emission ($i.e.$, outflow).
The origin of the elongated redshifted $^{13}$CO (2--1) emission to the west of the protostar
is not clear. As the elongation is along the outflow axis, this component may trace the part
of the outflow components located to the other side from the plane of the sky.

There must also be an underlying extended cloud component of the
$^{12}$CO and $^{13}$CO (2--1) emission. Interferometric observations of
such extended emission structures are challenging and often produce
sidelobe features, which are difficult to remove even with CLEAN.
The apparently noisier images in Figure \ref{fig:outflow}
likely reflect such an imperfectness of the deconvolution process.

From these outflow and previous results of the protostellar envelope
as seen in the C$^{18}$O (1--0) emission, the position angle of the
major axis of the protostellar envelope is regarded as P.A.=154$\degr$.

\subsection{C$^{18}$O (2--1), SO (6$_5$--5$_4$), and $^{13}$CO (3--2)}

\begin{figure}[ht!]
\figurenum{3}
\epsscale{1}
\begin{center}
\includegraphics[scale=0.5,angle=-90]{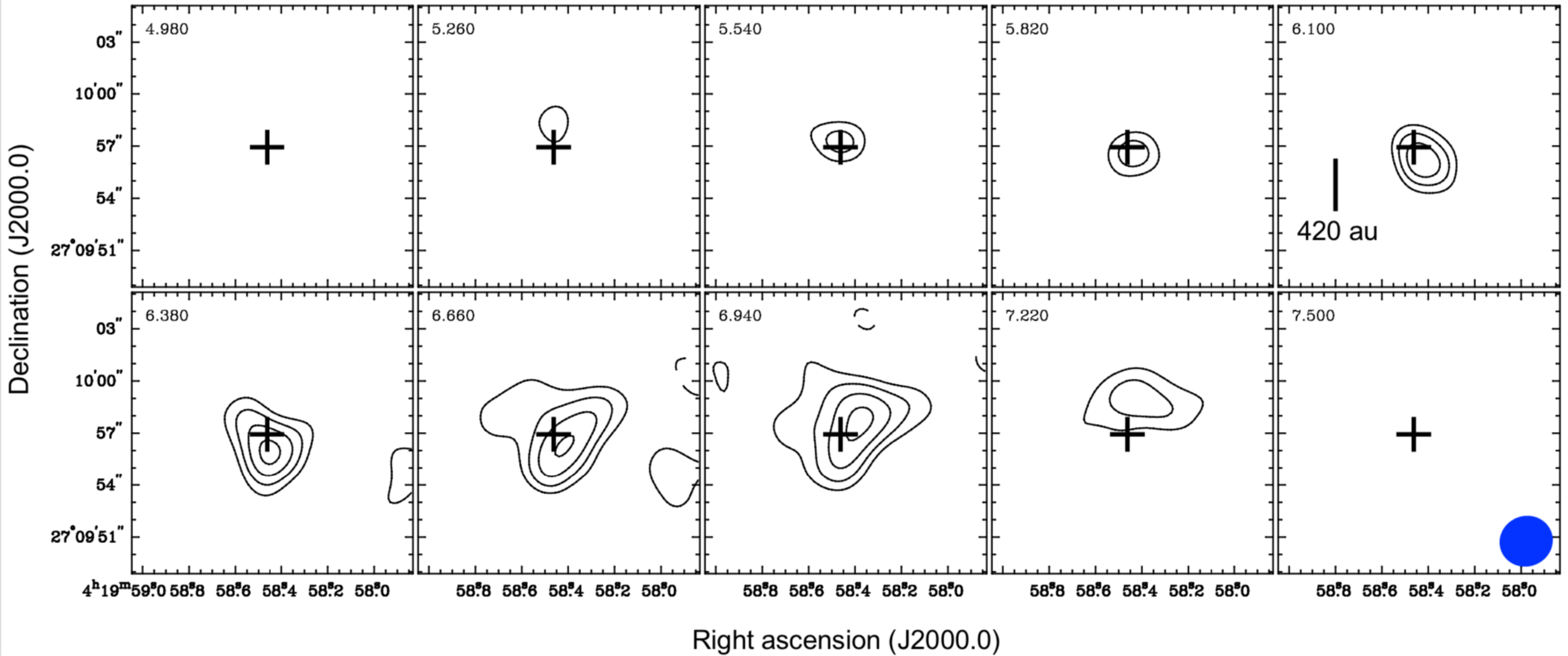}
\end{center}
\caption{Velocity channel maps of the C$^{18}$O (2--1) emission in I04169.
Contour levels are
3$\sigma$, 4$\sigma$, 5$\sigma$, 6$\sigma$ 
(1$\sigma$ = 0.099 Jy beam$^{-1}$). Crosses show the protostellar position, and
a filled ellipse at the bottom-right corner the synthesized beam
(3$\farcs$09$\times$2$\farcs$90; P.A. = -69$\degr$).
\label{fig:c18och}}
\end{figure}

Figure \ref{fig:c18och} shows velocity channel maps of the C$^{18}$O (2--1) emission
in I04169. At $V_{LSR}$ = 5.26 -- 5.54 km s$^{-1}$, there is an unresolved,
weak C$^{18}$O (2--1) component to the north of the protostar.
In the lower blueshifted velocities
(5.82 -- 6.66 km s$^{-1}$), the C$^{18}$O (2--1) emission
is located to the southwest of the protostar. In the redshifted velocities
(6.94 -- 7.22 km s$^{-1}$) the emission peak is shifted to the northwest and then north.
Thus, in the low-velocity C$^{18}$O (2--1) emission
the northern part is redshifted and the southern part blueshifted,
which is consistent with the velocity
gradient of the $r \sim$1000-au scale protostellar envelope \cite{oha97}.
The outermost extent of these low-velocity C$^{18}$O (2--1) emission is $r \sim$400 au.
Because of the correlator setting
of the SMA archival data, the C$^{18}$O (2--1) emission in the higher redshifted velocities
was not observed.
We have checked the phase and amplitude behaviors
in the C$^{18}$O (2--1) visibility spectra at the chunk edge.
The behaviors appear normal. Specifically,
the C$^{18}$O visibility spectrum taken with the compact configuration
shows a systematic phase trend changing from $\sim$+20$\degr$ to -20$\degr$
from channel 500 to 511, with an approximate rms of $\sim$10$\degr$.
This systematic phase trend reflects the observed 
north (red) to south (blue) velocity gradient. On the other hand,
the C$^{18}$O visibility spectrum taken with the extended configuration,
which is significantly down-weighted in the imaging process (see section 2),
is undetected. We consider that there is no significant edge effect of the SMA chunk,
which affects the observed C$^{18}$O (2--1) velocity gradient.

\begin{figure}[ht!]
\figurenum{4}
\epsscale{1}
\begin{center}
\includegraphics[scale=0.5,angle=-90]{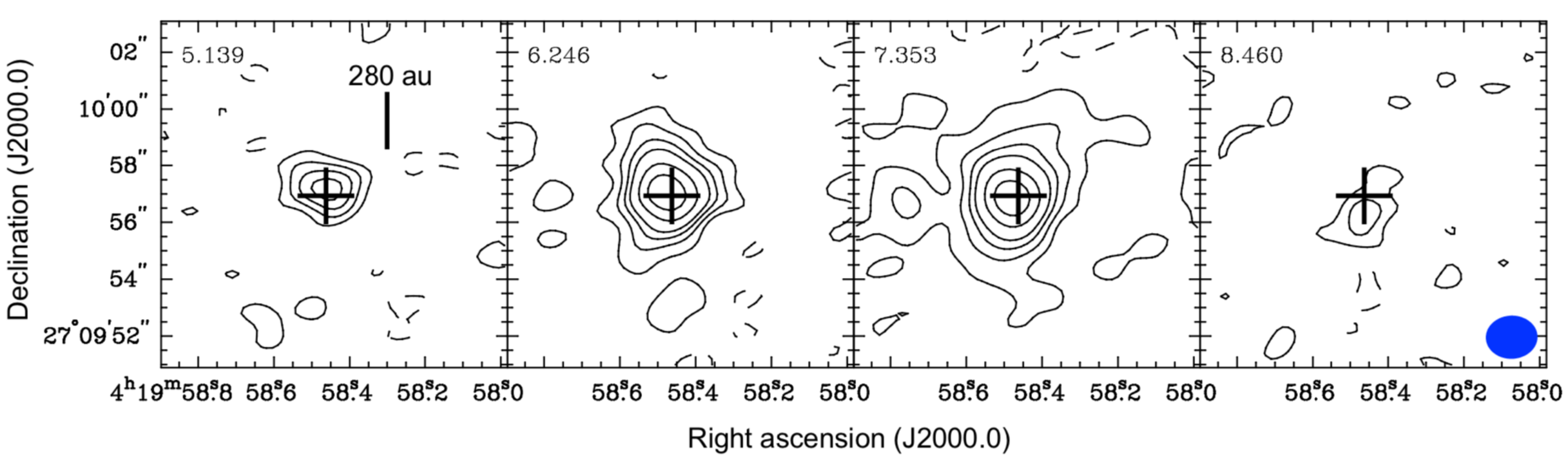}
\end{center}
\caption{Velocity channel maps of the SO (6$_5$--5$_4$) emission in I04169.
Contour levels are 2$\sigma$, 4$\sigma$, 6$\sigma$, 8$\sigma$, and then in steps
of 4$\sigma$ (1$\sigma$ = 0.027 Jy beam$^{-1}$).
Crosses show the protostellar position, and
a filled ellipse at the bottom-right corner the synthesized beam
(1$\farcs$82$\times$1$\farcs$52; P.A. = -88$\degr$).
\label{fig:soch}}
\end{figure}

Figure \ref{fig:soch} presents velocity channel maps of the SO (6$_{5}$--5$_{4}$) emission
in I04169. The SO emission is intense, and without any $uv$ tapering sufficient SO flux
densities are recovered within a finer beam size (1$\farcs$82$\times$1$\farcs$52) than
that of the C$^{18}$O (2--1) image cube. The image area in Figure \ref{fig:soch}
is correspondingly smaller than that in Figure \ref{fig:c18och}.
The velocity resolution of the SO image cube is, on the other hand, a factor 4 worse than
that of the C$^{18}$O (2--1) image cube.
In the highly blueshifted ($V_{LSR}$=5.14 km s$^{-1}$) and redshifted (8.46 km s$^{-1}$) velocities,
compact SO emission is seen to the north and south of the protostar, respectively.
The outermost extent of the high-velocity SO emission is $r \sim$200 au.
The peaks of the low-velocity blueshifted (6.25 km s$^{-1}$) and redshifted (7.35 km s$^{-1}$)
SO emission are also located to the north and south.
The sense of this SO velocity gradient is opposite to that of the
C$^{18}$O (2--1) emission and the $r \sim$1000-au scale protostellar envelope.
In the low-velocity range, there is also an emission feature
elongated approximately perpendicular to the outflow axis.

\begin{figure}[ht!]
\figurenum{5}
\epsscale{1}
\begin{center}
\includegraphics[scale=0.5,angle=-90]{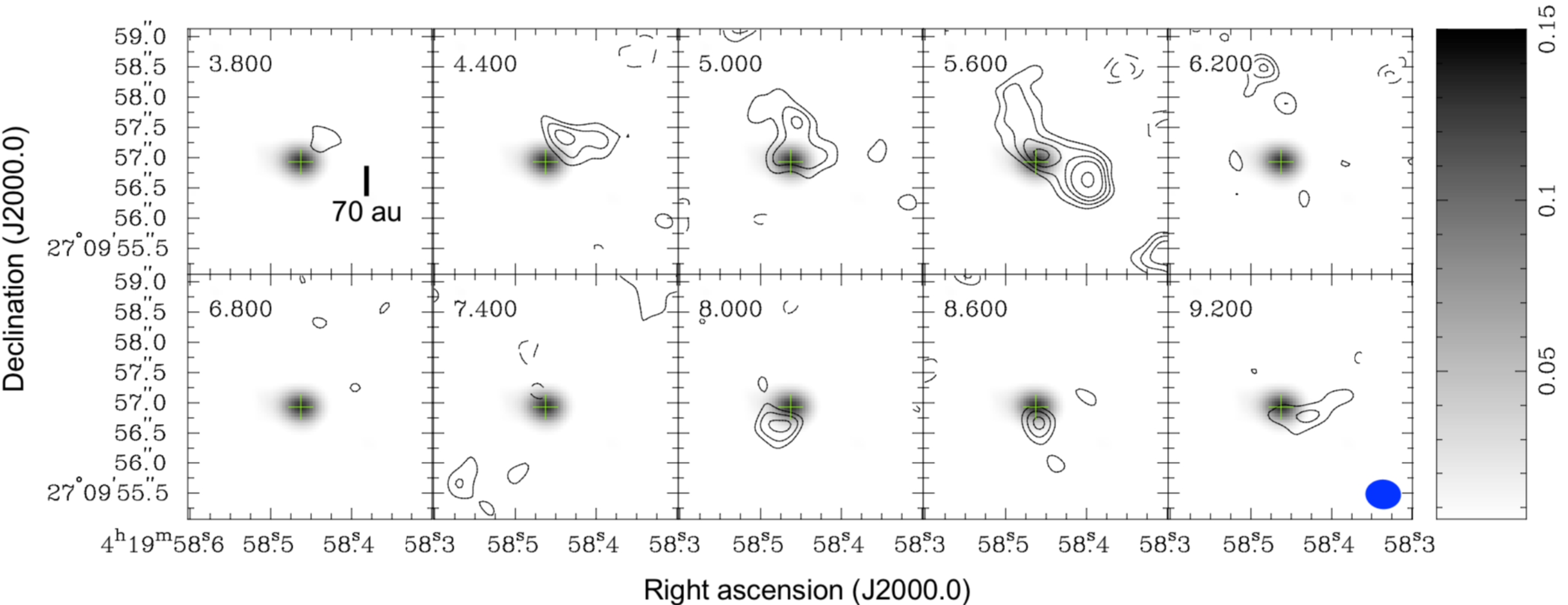}
\end{center}
\caption{Velocity channel maps of the $^{13}$CO (3--2) emission at every 0.6 km s$^{-1}$ bin
(contours), superposed on the 0.9-mm dust-continuum image with the Natural weighting
(gray), in I04169. Contour levels are
3$\sigma$, 4$\sigma$, 5$\sigma$, 6$\sigma$, and then in steps of 2$\sigma$
(1$\sigma$ = 69 mJy beam$^{-1}$). Crosses show the protostellar position.
A filled ellipse at the bottom-right corner shows the synthesized beam
(0$\farcs$59$\times$0$\farcs$49; P.A. = 86$\degr$).
\label{fig:13coch}}
\end{figure}

Figure \ref{fig:13coch} shows velocity channel maps of the $^{13}$CO (3--2) emission
in I04169. The higher angular-resolution ($\sim$0$\farcs$5) image cube reveals
a velocity structure of molecular gas in the vicinity ($r \sim$100 au) of the protostar.
Note that the angular resolution of this image cube is a factor $\sim$6 higher
than that of the C$^{18}$O (2--1) image cube described above.
For a clear presentation,
the image region of Figure \ref{fig:13coch} is set to be
a factor $\sim$4$^2$ smaller than that of the C$^{18}$O (2--1) image.
In the blueshifted velocity range ($V_{LSR}$=3.8--5.6 km s$^{-1}$),
there are $^{13}$CO (3--2) emission peaks located to the northwest of the protostar.
This blueshifted velocity range approximately matches that of the SO velocity channel
at $V_{LSR}$=5.14 km s$^{-1}$, where the SO emission is also located to the north
(Figure \ref{fig:soch}). There is also a weak C$^{18}$O (2--1) component seen to
the north at $V_{LSR}$ = 5.26 -- 5.54 km s$^{-1}$, consistent with the
$^{13}$CO (3--2) velocity range. Therefore, the northern highly blueshifted gas component
is likely seen in all the three tracers.
At $V_{LSR}$ = 5.6 km s$^{-1}$, another $^{13}$CO (3--2) emission component to the southwest
of the protostar is seen. As the location of this emission component is close to
the outflow axis and the southwestern side corresponds to the blueshifted outflow side
(Figure \ref{fig:outflow}), this emission component is likely to trace the outflow.
In the lower velocity range at $v_{sys}$ (= 6.8 km s$^{-1}$) $\pm$0.6 km s$^{-1}$,
the $^{13}$CO (3--2) emission appears to be severely suppressed. This is presumably due
to the presence of a foreground, intervening molecular cloud material,
which absorbs the $^{13}$CO (3--2) emission associated with the Class I object.
In these velocities, the low-velocity C$^{18}$O (2--1) emission exhibits
a south (blue) to north (red) velocity gradient as described above
(Figure \ref{fig:c18och}).
In the redshifted velocity range ($V_{LSR}$ = 8.0--9.2 km s$^{-1}$),
the $^{13}$CO (3--2) emission is located to the southeast and south of the protostar.
This redshifted velocity range is approximately consistent with that of the SO velocity channel
at $V_{LSR}$=8.46 km s$^{-1}$, where the SO emission is also located to the south
(Figure \ref{fig:soch}).
%
%

\begin{figure}[ht!]
\figurenum{6}
\epsscale{1}
\includegraphics[scale=0.6,angle=-90]{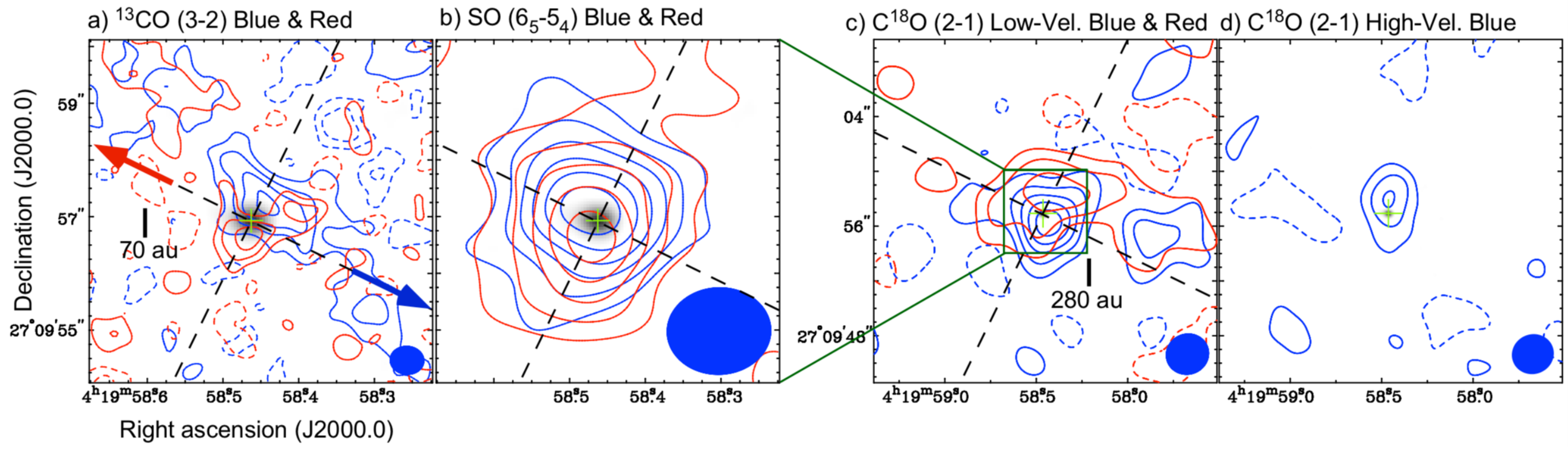}
\caption{a) Maps of the blueshifted (blue contours) and redshifted (red) $^{13}$CO (3--2) emission,
superposed on the 0.9-mm dust-continuum image (gray), in I04169.
The integrated velocity ranges for the blueshifted and redshifted emission are
$V_{LSR}$ = 3.7 -- 5.9 km s$^{-1}$ and 7.5 -- 9.3 km s$^{-1}$, respectively.
Contour levels are in steps of 2$\sigma$, where 1$\sigma$ noise levels
are 80 mJy beam$^{-1}$ km s$^{-1}$ and 72 mJy beam$^{-1}$ km s$^{-1}$ for the blueshifted
and redshifted emission, respectively. A cross shows the position of the protostar.
Dashed lines denote the major and minor axes of the protostellar envelope.
Blue and red arrows show the directions of the
blueshifted and redshifted outflows (Figure \ref{fig:outflow}).
A filled ellipse at the bottom-right corner shows the synthesized beam
(0$\farcs$59$\times$0$\farcs$49; P.A. = 86$\degr$).
b) Maps of the blueshifted and redshifted SO (6$_{5}$--5$_{4}$) emission in I04169.
The integrated velocity ranges for the blueshifted and redshifted emission are
$V_{LSR}$ = 4.59 -- 6.80 km s$^{-1}$ and 6.80 -- 9.01 km s$^{-1}$, respectively.
Contour levels are in steps of 3$\sigma$, where the 1$\sigma$ noise level is
42 mJy beam$^{-1}$ km s$^{-1}$.
A filled ellipse at the bottom-right corner shows the synthesized beam
(1$\farcs$82$\times$1$\farcs$52; P.A. = -88$\degr$).
c) Maps of the low-velocity blueshifted and redshifted C$^{18}$O (2--1) emission
superposed on the 0.9-mm dust-continuum image in I04169.
The integrated velocity ranges are $V_{LSR}$ = 5.68 -- 6.80 km s$^{-1}$ and 6.80 -- 7.64 km s$^{-1}$
for the blueshifted and redshifted emission, respectively.
Contour levels are in steps of 2$\sigma$, where 1$\sigma$ noise levels
are 55 mJy beam$^{-1}$ km s$^{-1}$ and 48 mJy beam$^{-1}$ km s$^{-1}$ for the blueshifted
and redshifted emission, respectively.
A filled ellipse at the bottom-right corner shows the synthesized beam
(3$\farcs$09$\times$2$\farcs$90; P.A. = -69$\degr$).
d) Map of the high-velocity blueshifted ($V_{LSR}$ = 4.28 -- 5.68 km s$^{-1}$)
C$^{18}$O (2--1) emission.
Contour levels are in steps of 2$\sigma$ (1$\sigma$ = 62 mJy beam$^{-1}$ km s$^{-1}$).
\label{fig:velall}}
\end{figure}

Figure \ref{fig:velall} compares the detected velocity features
in the different molecular tracers at different spatial scales. In the $^{13}$CO (3--2) emission,
the high-velocity blueshifted and redshifted components are located to the northwest
and southeast of the protostar, respectively (Figure \ref{fig:velall}a).
The outermost extent of the $^{13}$CO (3--2)
emission is $r \sim$100 au.
The apparently noisier image of the $^{13}$CO (3--2) map is
presumably due to the contamination from an extended, foreground absorbing material.
In the $^{13}$CO (3--2) velocity channel maps (Figure \ref{fig:13coch}),
the beam size is 0$\farcs$589 $\times$ 0$\farcs$490 $\sim$0$\farcs$537 (geometrical mean),
and the peak S/N are 5.5$\sigma$, 5.2$\sigma$, 6.3$\sigma$, 5.3$\sigma$,
and 5.3$\sigma$ at $V_{LSR}$ = 4.4 km s$^{-1}$, 5.0 km s$^{-1}$, 5.6 km s$^{-1}$,
8.0 km s$^{-1}$, and 8.6 km s$^{-1}$, respectively.
Thus the positional accuracy at each velocity channel should be better
than $\sim$0$\farcs$537 / 5.2$\sigma$ $\sim$0$\farcs$103.
The angular separation between the blueshifted and redshifted emission is
$\sim$0$\farcs$65 (Figure \ref{fig:velall}a),
and thus this separation between the blueshifted and redshifted velocities
($i.e.$, velocity gradient) is detected at least above
$\sim$0$\farcs$65 / 0$\farcs$103 $\sim$6.3$\sigma$.
If the blueshifted and redshifted emission are integrated
as shown in Figure \ref{fig:velall}a, the peak S/N of the blueshifted and
redshifted emission are 8.5$\sigma$ and 7.4$\sigma$, respectively. Thus
the significance of the velocity gradient is raised to
$\sim$0$\farcs$65 / 0$\farcs$537 / 7.4$\sigma$ $\sim$ 9.0$\sigma$. 
These estimates show that the detected northwest (blue) to southeast (red)
velocity gradient in the $^{13}$CO (3--2) emission is statistically robust. 
Furthermore, the direction of the velocity gradient approximately matches with
the major axis of the protostellar envelope, and the axis
orthogonal to the outflow direction.
The peak locations of the integrated blueshifted and redshifted
SO (6$_{5}$--5$_{4}$) emission appear to match with those of the $^{13}$CO (3--2) emission,
although the spatial resolution of the SO map is factor $\sim$3 worse
(Figure \ref{fig:velall}b).

On the other hand, in the lower-velocity range
where the $^{13}$CO (3--2) emission is significantly suppressed,
the blueshifted and redshifted C$^{18}$O (2--1) emission are located to the south and north
of the protostar, respectively, with the outermost extent of $r \sim$400 au (Figure \ref{fig:velall}c).
The sign of the C$^{18}$O (2--1) velocity gradient is opposite to that
of the inner $^{13}$CO (3--2) and SO (6$_{5}$--5$_{4}$) emission.
Furthermore, the direction of the axis connecting the peaks of the blueshifted
and redshifted C$^{18}$O (2--1) emission is along north to south, rather than northwest to southeast.
This indicates that there is also a slight velocity gradient in the
C$^{18}$O (2--1) emission along the minor axis, and the northeastern part is slightly more
redshifted and southwestern part more blueshifted.
There is also a compact, high-velocity blueshifted C$^{18}$O (2--1) emission located to the north
of the protostar (Figure \ref{fig:velall}d). As the spatial location and the velocity range
of this C$^{18}$O (2--1) emission
component is consistent with those of the blueshifted $^{13}$CO (3--2) emission, the origin of
the high-velocity blueshifted C$^{18}$O (2--1) emission is likely the same as that of the
$^{13}$CO (3--2) emission.

\begin{figure}[ht!]
\figurenum{7}
\epsscale{1}
\begin{center}
\includegraphics[scale=0.6,angle=-90]{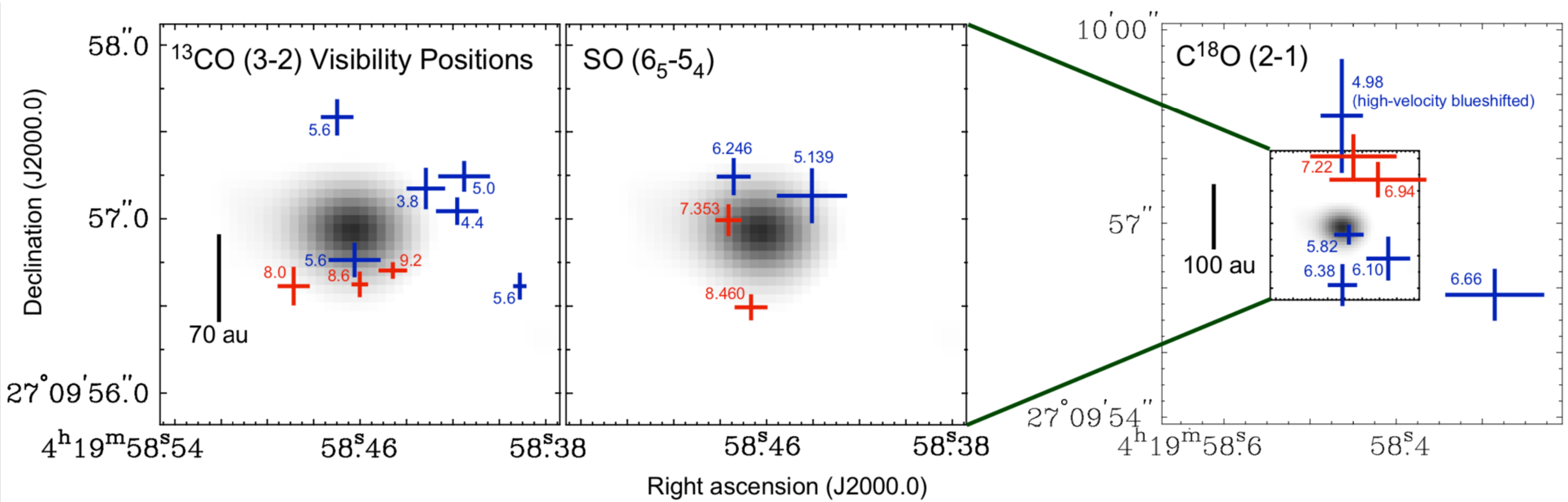}
\end{center}
\caption{Plots of the peak position
of the $^{13}$CO (3--2) (left panel), SO (6$_5$--5$_4$) (middle),
and C$^{18}$O (2--1) emission (right) at each velocity,
as derived from direct fitting
of 2-dimensional Gaussian functions to the visibilities. 
Blue and red crosses denote the peak positions of
the blueshifted and redshifted emission, respectively,
and the number next to each cross shows
the LSR velocity. Lengths of the crosses denote
the fitting errors. Background
gray scales show the 0.9-mm continuum image
(same as Figure \ref{fig:contall}a).
\label{fig:vispos}}
\end{figure}

To further verify the presence of the opposite velocity gradients
between the $^{13}$CO, SO, and the C$^{18}$O emission, we have also fit
2-dimensional Gaussians to the observed visibility data
and derived the emission peak positions at different velocity bins.
The velocity bins are chosen to match with those of the relevant
velocity channel
maps (Figures \ref{fig:c18och}, \ref{fig:soch}, \ref{fig:13coch}),
except for the high-velocity blueshifted C$^{18}$O emission
(Figure \ref{fig:velall}d). A single 2-dimensional Gaussian is
fitted to the visibility at each velocity bin, except for the
$^{13}$CO (3--2) emission at $V_{LSR}$ = 5.6 km s$^{-1}$ where three
Gaussians are adopted. The fitting results are summarized
in Figure \ref{fig:vispos}. In the $^{13}$CO (3--2) visibility fitting,
the blueshifted positions are clustered to the north and northwest
while the redshifted positions to the south. This trend reflects
the well-separated blueshifted and redshifted emission as shown in
Figure \ref{fig:velall}a. The two blue- and redshifted points of
the SO emission are also well separated and located to the north
and south, respectively. On the other hand, in the case of the
C$^{18}$O emission two redshifted points
are located to the north while the four blueshifted points
to the south. The exception is the high-velocity blueshifted
C$^{18}$O emission located to the north.
These visibility analyses demonstrate that the velocity structures
and the flip of the velocity gradient between the
$^{13}$CO, SO, and the C$^{18}$O emission identified in the images
(Figure \ref{fig:velall}) are also present in
the visibility domain, and are unlikely due to the interferometric
imaging effect.

Table \ref{sumvel} summarizes the detected velocity gradients
along the major axis.

\begin{deluxetable}{cccc}
\tabletypesize{\scriptsize}
\tablecaption{Summary of the Velocity Gradients around I04169 along the Major Axis\label{sumvel}}
\tablewidth{0pt}
\tablehead{\colhead{Molecular Line} &\colhead{Radius} &\colhead{Velocity Gradient} &\colhead{References}\\
\colhead{} &\colhead{(au)} &\colhead{(NW -- SE)} &\colhead{}}
\startdata
C$^{18}$O (1--0)      &$\sim$1000 &Red -- Blue   &Ohashi et al. (1997) \\
C$^{18}$O (2--1)      &$\sim$400  &Red -- Blue   &This work \\
SO (6$_{5}$--5$_{4}$) &$\sim$200  &Blue -- Red   &This work \\
$^{13}$CO (3--2)      &$\sim$100  &Blue -- Red   &This work \\
\enddata
\end{deluxetable}

\subsection{Position - Velocity Diagrams}

\begin{figure}[ht!]
\figurenum{8}
\epsscale{1}
\begin{center}
\includegraphics[scale=0.7,angle=0]{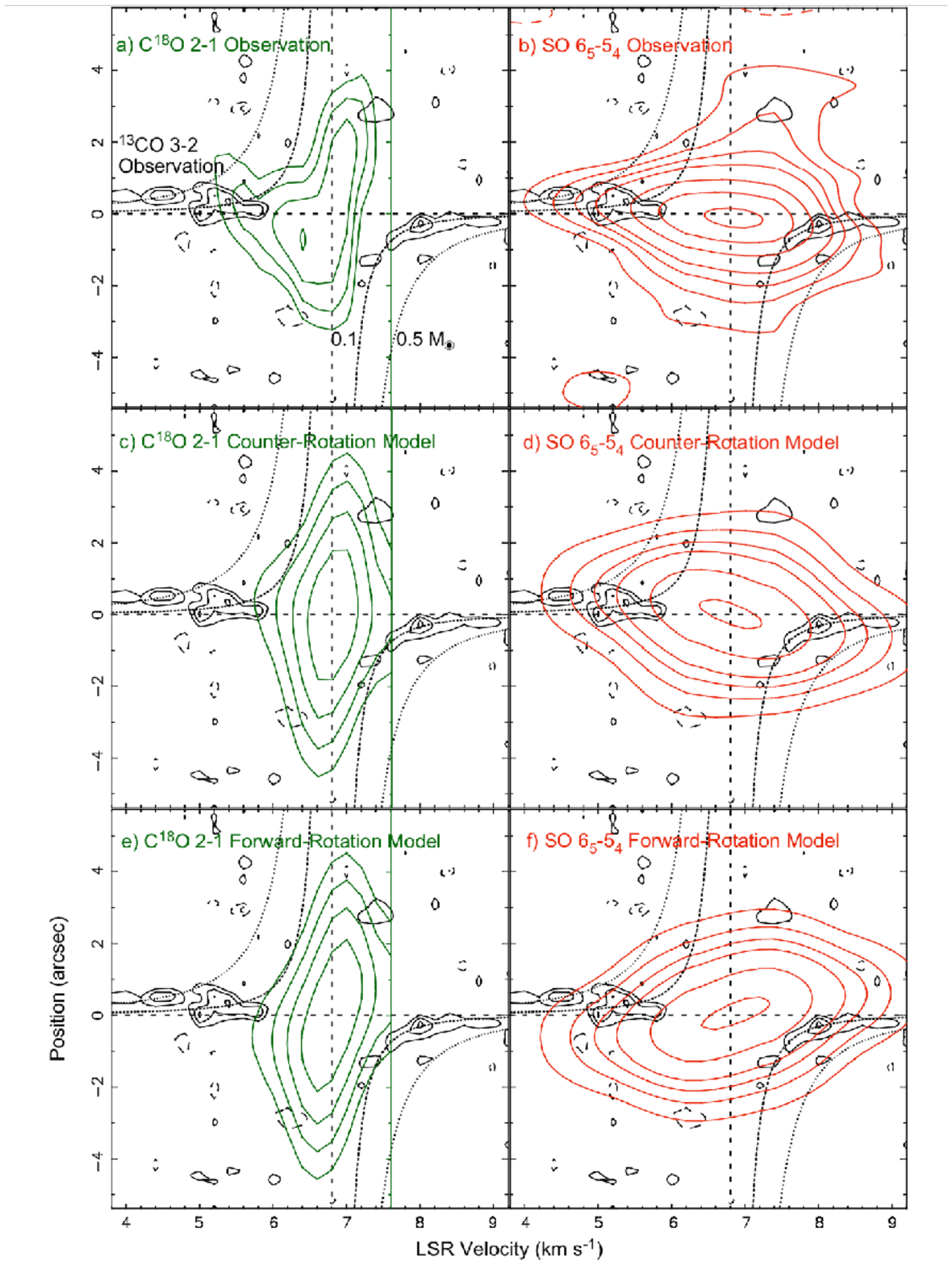}
\end{center}
\caption{a) Observed Position - Velocity (P-V) diagrams of the $^{13}$CO (3--2)
(black contour) and C$^{18}$O (2--1) emission (green)
in I04169 along the major axis (P.A. = 154$\degr$) passing through
the protostellar position.
Contour levels of the $^{13}$CO P-V
start from 2$\sigma$ in steps of 1$\sigma$ (1$\sigma$ = 0.12 Jy beam$^{-1}$),
and those of the C$^{18}$O P-V from 3$\sigma$ in steps of 1$\sigma$
(1$\sigma$ = 0.099 Jy beam$^{-1}$).
Vertical and horizontal dashed lines denote the systemic velocity of 6.8 km s$^{-1}$ and
the protostellar position, respectively. Dashed curves show Keplerian rotation curves
with $i$ = 60$\degr$ and the central protostellar masses of 0.1 $M_{\odot}$ and 0.5 $M_{\odot}$
as labeled. The rightmost green vertical line delineates
the spectral range covered with the present SMA C$^{18}$O (2--1) data.
b) Observed P-V diagram of the SO (6$_5$--5$_4$) emission in I04169 along the major axis
passing through the protostellar position (red contours),
overlaid on the P-V diagram of the $^{13}$CO (3--2) emission (black contours).
Contour levels of the SO emission are 2$\sigma$, 4$\sigma$, 6$\sigma$, 8$\sigma$,
and then in steps of 4$\sigma$ (1$\sigma$ = 0.027 Jy beam$^{-1}$).
c) P-V diagram of the C$^{18}$O (2--1) emission along the major axis obtained from
the counter-rotating model. Contour levels are the same as those
of the observed P-V diagram.
d) P-V diagram of the SO (6$_5$--5$_4$) emission along the major axis obtained from
the same counter-rotating model as that of the C$^{18}$O (2--1) model.
Contour levels are the same as those of the observed P-V diagram.
e) and f) P-V diagrams of the C$^{18}$O (2--1) and SO (6$_5$--5$_4$) emission,
respectively, obtained from the forward-rotating model.
\label{fig:pvmaj}}
\end{figure}

Figures \ref{fig:pvmaj}a and b show Position - Velocity (P-V) diagrams
of the C$^{18}$O (2--1) (green contours) and SO (6$_5$--5$_4$) emission (red)
passing through the protostellar position along the major axis (NW - SE),
superposed on the $^{13}$CO (3--2) P-V diagram (black).
The cut line of the P-V diagrams is shown in Figure \ref{fig:velall}.
The $^{13}$CO P-V diagram along the major axis
shows that the blue and redshifted emission are spatially separated
with respect to the protostellar position.
The spatial and velocity locations of the $^{13}$CO emission
overlap with the Keplerian rotation curve
around the central stellar mass of 0.1 $M_{\odot}$
(see dashed curves in Figure \ref{fig:pvmaj}),
although the limited sensitivity and spatial dynamical range
of the present SMA data prevent us from deriving the radial profile
of the gas motion.
Within the velocity range where the $^{13}$CO emission is
significantly suppressed,
the C$^{18}$O emission shows that the northwestern part
is redshifted and southeastern part blueshifted.
The sign of this C$^{18}$O velocity gradient is opposite to that of the $^{13}$CO emission.
In the higher blueshifted velocity there is an additional C$^{18}$O emission component,
which appears to connect to the blueshifted $^{13}$CO emission.

Figure \ref{fig:pvmaj}b shows that the high-velocity ($\gtrsim$ $V_{sys} \pm$0.8 km s$^{-1}$)
SO emission appears to be closely correlated with the $^{13}$CO emission, and
the high-velocity blueshifted SO emission is located to the northwest and the redshifted
emission southeast.
On the other hand, in the lower-velocity range ($\lesssim$ $V_{sys} \pm$0.8 km s$^{-1}$)
the extended SO emission appears to trace a similar velocity feature to that traced by the
C$^{18}$O emission.

\begin{figure}[ht!]
\figurenum{9}
\epsscale{1}
\begin{center}
\includegraphics[scale=0.5,angle=-90]{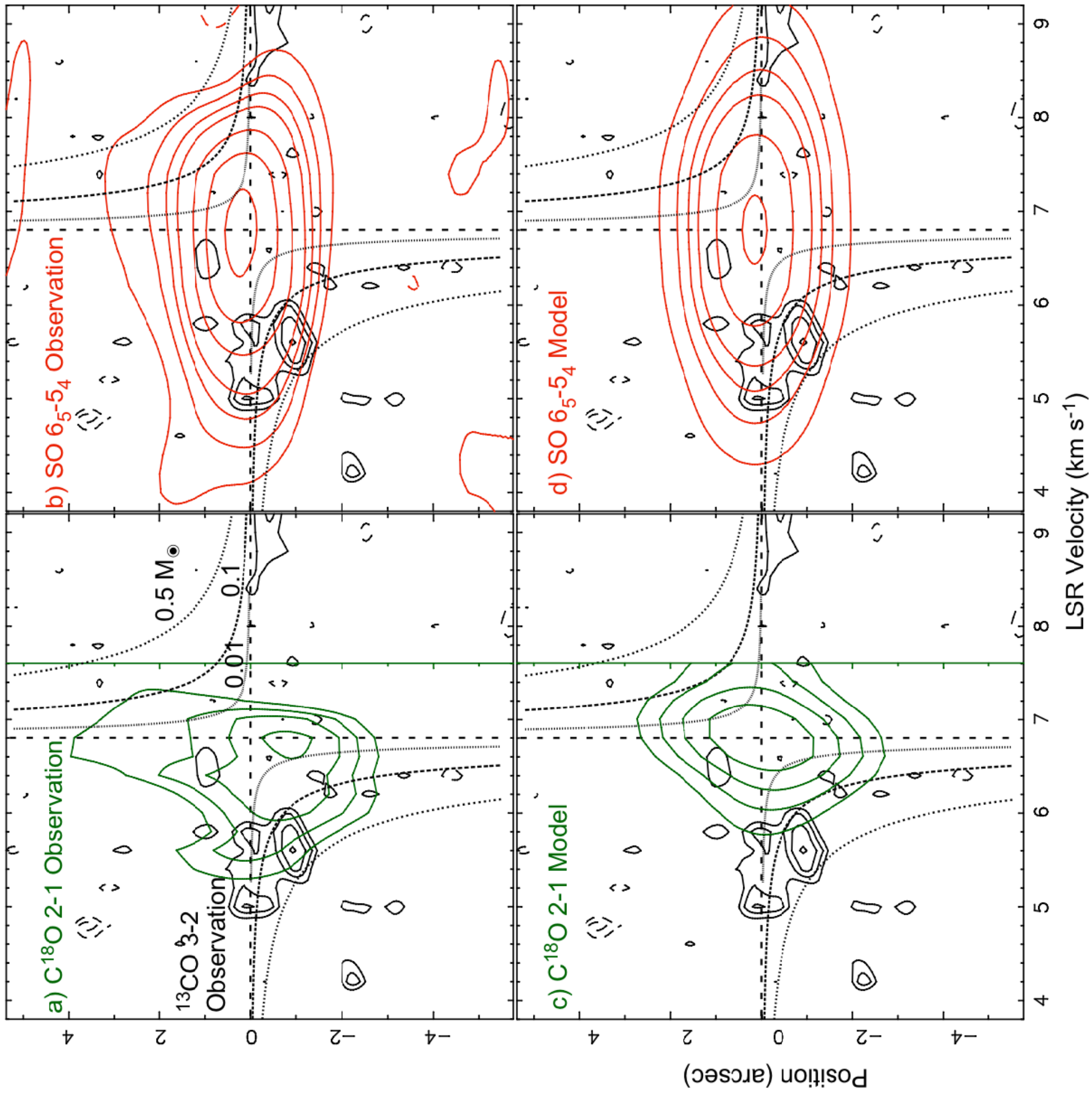}
\end{center}
\caption{a) Observed P-V diagram of the C$^{18}$O (2--1) emission
in I04169 along the minor axis (P.A. = 64$\degr$) passing through
the protostellar position (green contours),
overlaid on the P-V diagram of the $^{13}$CO (3--2) emission (black).
Contour levels of the C$^{18}$O emission start from 3$\sigma$ in steps of 1$\sigma$
(1$\sigma$ = 0.099 Jy beam$^{-1}$), and those of the $^{13}$CO (3--2) emission
from 2$\sigma$ in steps of 1$\sigma$ (1$\sigma$ = 0.12 Jy beam$^{-1}$).
Vertical and horizontal dashed lines denote the systemic velocity of 6.8 km s$^{-1}$ and
the protostellar position, respectively. Dashed curves show free-fall curves
with $i$ = 60$\degr$ and the central protostellar masses of 0.01 $M_{\odot}$,
0.1 $M_{\odot}$ and 0.5 $M_{\odot}$ as labeled.
The rightmost green vertical line delineates
the spectral range covered with the present SMA C$^{18}$O (2--1) data.
b) Observed P-V diagram of the SO (6$_{5}$--5$_{4}$) emission
in I04169 along the minor axis (P.A. = 64$\degr$) passing through
the protostellar position (red contours),
overlaid on the P-V diagram of the $^{13}$CO (3--2) emission (black).
Contour levels of the SO emission are 2$\sigma$, 4$\sigma$, 6$\sigma$, 8$\sigma$,
and then in steps of 4$\sigma$ (1$\sigma$ = 0.027 Jy beam$^{-1}$).
c) and d) Model P-V diagrams of the C$^{18}$O (2--1) and SO (6$_5$--5$_4$) emission,
respectively. Contour levels are the same as those of the corresponding observed P-V diagrams.
\label{fig:pvmin}}
\end{figure}

Figures \ref{fig:pvmin}a and b compare the P-V diagrams of the $^{13}$CO (3--2) (black contours),
C$^{18}$O (2--1) (green), and the SO (6$_{5}$--5$_{4}$) (red) emission along the minor axis.
As described above, the C$^{18}$O (2--1) emission shows
a slight velocity gradient along the minor axis too, where
the southwestern part is blueshifted and the
northeastern part redshifted (Figure \ref{fig:pvmin}a).
Since the associated molecular outflow is blue- and redshifted
to the southwest and northeast (Figure \ref{fig:outflow}), the southwestern and northeastern
parts of the flattened
protostellar envelope perpendicular to the outflow must be on the far- and near side.
Thus, the C$^{18}$O emission exhibits the blueshifted emission
on the far side and redshifted emission on the near side, which can be interpreted as an infalling gas motion
on the flattened envelope. Based on this concept, free-fall curves on the mid-plane with
the central protostellar mass of 0.5, 0.1, and 0.01 $M_{\odot}$ are drawn in the P-Vs.
Whereas the limited spatial resolution and sensitivity of the C$^{18}$O image cube
prevent us from making conclusive
discussion, the C$^{18}$O velocity gradient along the minor axis seems to be
better traced with the free-fall curve with the central mass lower than 0.1 $M_{\odot}$.
The P-V diagram of the SO emission along the minor axis does not show a clear velocity gradient,
in contrast with the P-V diagram along the major axis.

\subsection{Line Profiles}

\begin{figure}[ht!]
\figurenum{10}
\epsscale{1}
\begin{center}
\includegraphics[scale=0.4,angle=0]{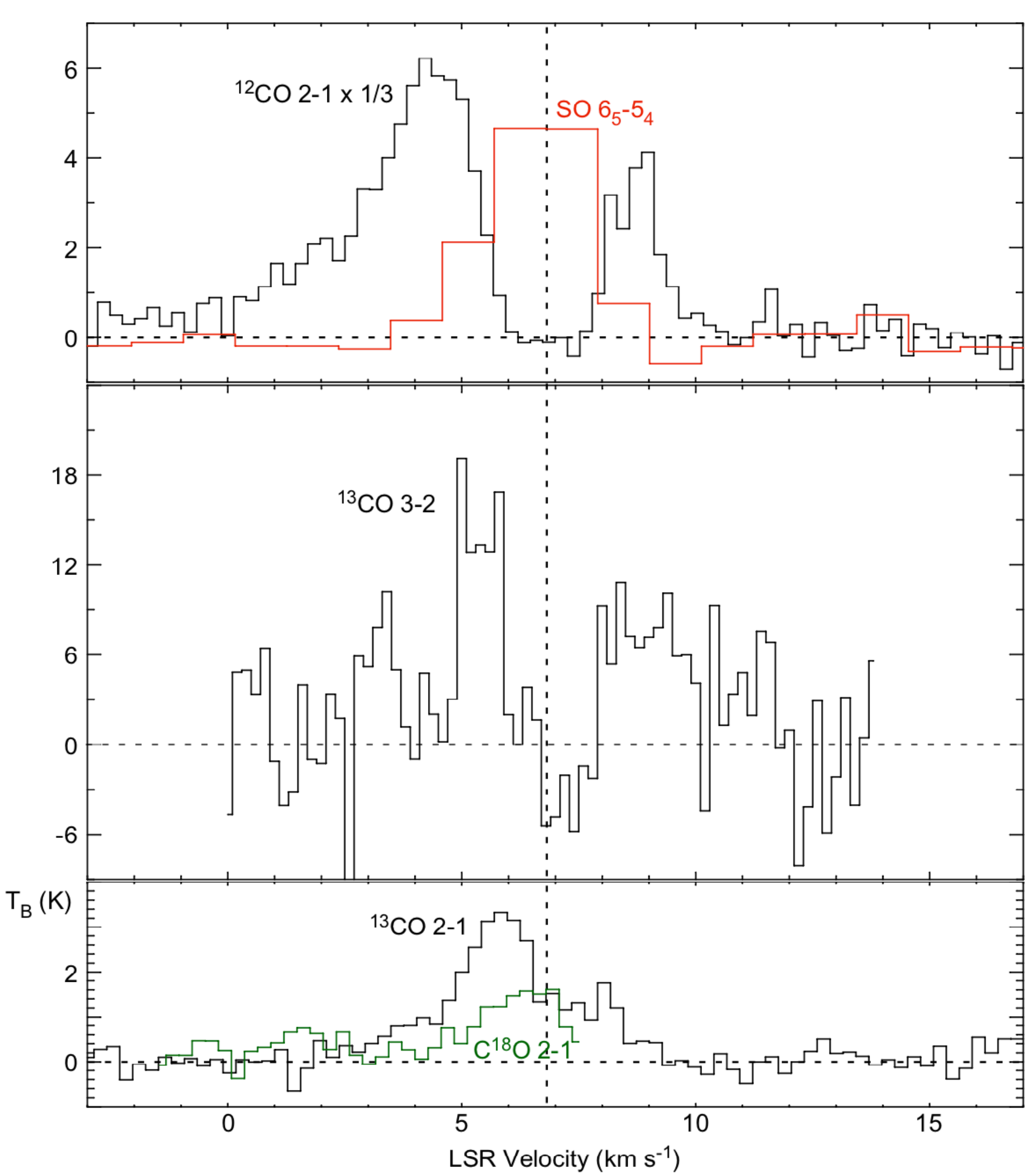}
\end{center}
\caption{SMA spectra toward the protostellar position of I04169.
For direct comparison, the beam sizes of the $^{12}$CO (2--1)
and $^{13}$CO (2--1) spectra are smoothed to those of the SO (6$_5$--5$_4$)
and C$^{18}$O (2--1) spectra, respectively (see Table \ref{lineobs}).
Vertical and horizontal dashed lines denote the systemic velocity of
6.8 km s$^{-1}$ and the spectral baseline, respectively.
\label{fig:spec}}
\end{figure}

Figure \ref{fig:spec} shows the SMA line profiles toward the protostellar position of I04169.
The brightness temperature of the $^{12}$CO (2--1) line
is scaled by a factor of 1/3 for more direct comparison between the $^{12}$CO (2--1)
and SO (6$_5$--5$_4$) line. The $^{12}$CO (2--1) line profile
exhibits a blueshifted high-velocity wing until $V_{LSR} \lesssim$0.0 km s$^{-1}$,
consistent with our
interpretation that the $^{12}$CO (2--1) emission traces the molecular outflow.
In the redshifted part there is only a marginal hint of the wing, but the $U$-shaped
spatial distribution of the redshifted $^{12}$CO (2--1) emission
(see Figure \ref{fig:outflow}a) suggests that
the origin of the redshifted $^{12}$CO (2--1) emission is also the molecular outflow.
The SO (6$_{5}$--5$_{4}$) line does not show such a wing component, but a single peak
at around the systemic velocity, where the $^{12}$CO (2--1) emission is largely suppressed.
While the signal-to-noise ratio is limited,
no clear blueshifted wing is identified in the the $^{13}$CO (3--2) spectrum.
Around the systemic velocity the $^{13}$CO (3--2) emission is significantly suppressed,
presumably due to the presence of a foreground absorbing material.
Such a foreground absorber can also explain
the absorption features in the $^{12}$CO (2--1) and $^{13}$CO (2--1) spectra
at the same velocities.
The peak velocity of the blueshifted $^{13}$CO (3--2) line ($V_{LSR} \sim$5.5 km s$^{-1}$)
is located where the $^{12}$CO (2--1) emission becomes deficient.
Comparison between the $^{12}$CO (2--1) and $^{13}$CO (3--2)
velocity channel maps at the same angular resolution
shows that the peak positions and the systematic velocity structures
are different between the $^{12}$CO (2--1) and $^{13}$CO (3--2) emission.
These results imply that the nature of the $^{13}$CO (3--2) and SO (6$_{5}$--5$_{4}$)
emission is distinct from that of the $^{12}$CO (2--1) emission ($i.e.$, outflow), and that the
$^{13}$CO (3--2) and SO (6$_{5}$--5$_{4}$) emission most likely trace a circumstellar
gas component.

From the two isotropic lines of the $^{13}$CO (2--1) and C$^{18}$O (2--1) emission,
the excitation temperature and the optical depths of the molecular lines can be estimated
by solving the following simultaneous equations,
on the assumption that both the $^{13}$CO (2--1) and C$^{18}$O (2--1) emission
trace the same gas component and are in the LTE condition,
\begin{equation}
T_B^{13} = (J_{\nu_{13}}(T_{\rm ex})-J_{\nu_{13}}(T_{bg}))(1-\exp(-f_{r} \tau_{18})),
\end{equation}
\begin{equation}
T_B^{18} = (J_{\nu_{18}}(T_{\rm ex})-J_{\nu_{18}}(T_{bg}))(1-\exp(-\tau_{18})),
\end{equation}
where $T_B^{X}$, $\nu_{X}$, $\tau_{X}$ ($X$=13 or 18) are the brightness temperature,
line frequency, and the optical depth of the $^{13}$CO (2--1) or C$^{18}$O (2--1) emission.
$T_{\rm ex}$ is the excitation temperature, which is common between the 
$^{13}$CO (2--1) and C$^{18}$O (2--1) emission on the assumption of the LTE condition,
and $T_{bg}$ is the background temperature (=2.725 K). $f_{r}$ is the ratio of the
$^{13}$CO and C$^{18}$O molecular abundance, and is adopted to be $f_{r}$ = 7.3 \cite{wr94}.
Around the systemic velocity, the $^{13}$CO (2--1) brightness temperature becomes below
the C$^{18}$O (2--1) brightness temperature, suggesting that the $^{13}$CO (2--1) emission
also suffers from the effect of the foreground absorbing material.
In these velocity regions it is not
possible to estimate $T_{\rm ex}$ and $\tau_{18}$ from the above equations. Thus,
$T_{\rm ex}$ and $\tau_{18}$ are estimated at bluer velocities, and at $V_{LSR}$ = 5.8,
6.1, and 6.4 km s$^{-1}$ $T_{\rm ex}$ is estimated to be 7.8, 7.5, and 6.9 K,
and $\tau_{18}$ 0.43, 0.61, and 0.87, respectively.
We note, however, that there is likely a significant contamination from the outflow component
to the $^{13}$CO (2--1) emission (Figure \ref{fig:outflow}), which makes this simple one-zone
LTE analysis infeasible. On the other hand, the observed peak brightness temperature of
the C$^{18}$O (2--1) emission ($\lesssim$1.6 K) is well below the anticipated gas temperature
of the protostellar envelope ($\sim$ 20 K) \cite{aso15,har15}, which suggests that
the C$^{18}$O (2--1) emission is likely optically thin.

The SO (6$_{5}$--5$_{4}$) image cube is convolved to have the same angular resolution as that of the
$^{13}$CO (2--1) and C$^{18}$O (2--1) image cubes shown in Figure \ref{fig:spec},
and $T_{\rm ex}$ of the SO (6$_{5}$--5$_{4}$) line is assumed to be the same
as derived above (=7.8 K).
Then, the optical depth of the SO (6$_{5}$--5$_{4}$) emission is estimated to be 0.26, 1.10, 1.15, and 0.13
at $V_{LSR}$ = 5.1, 6.2, 7.4, and 8.5 km s$^{-1}$, respectively. The estimated SO (6$_{5}$--5$_{4}$)
optical depth is likely an upper limit, as the SO (6$_{5}$--5$_{4}$) emission
traces the inner, denser molecular gas
than that traced by the C$^{18}$O (2--1) emission and the excitation temperature
of the SO (6$_{5}$--5$_{4}$) emission
is likely higher than that of the C$^{18}$O (2--1) emission. These results indicate
that the SO (6$_{5}$--5$_{4}$) emission is also optically thin or marginally optically thick.
Therefore, the effects of the optical depths on the gas kinematics
traced by these two lines are likely negligible.

\section{Analysis}

The SMA image cubes in a set of the molecular emission toward I04169
exhibit distinct velocity structures at different spatial scales.
To investigate the nature of these velocity structures,
we constructed geometrically-thin envelope-disk models in the
C$^{18}$O (2--1) and SO (6$_{5}$--5$_{4}$) emission.
As the purpose of our model is to reproduce the observed gas motions in the inner
$\lesssim$500 au scale around the protostar, geometrically-thin
assumption is likely valid.
Even at a larger scale, the C$^{18}$O (1--0) emission shows a
2200 au $\times$ 1100 au scale elongated envelope around the protostar,
and the major axis of the envelope is perpendicular to the associated
outflow axis \cite{oha97}.
The outflow is well developed with a wide opening angle ($>$40$\degr$;
Figure \ref{fig:outflow}a) at around the apex. These observational results
imply that dense gas distribution within $r \sim$1000 au from the protostar
is likely flattened. Furthermore, our modeling of protostellar envelopes
and disks around the other protostellar sources as observed with ALMA reveals
that geometrically-thin envelope / disk models reproduce the observed features
reasonably well ($e.g.$, Aso et al. 2015; Yen et al. 2017; Takakuwa et al. 2017).
The formation of so-called pseudo-disks is also expected from several
MHD simulations \cite{ma11a,har15,tom15}.

The protostellar envelope around I04169 as seen in the C$^{18}$O (1--0) emission
is approximated to be a 2-dimensional Gaussian with the FWHM sizes along
the major $\times$ minor axes of
16$\arcsec$$\times$8$\arcsec$ and the position angle of 154$\degr$ \cite{oha97}.
Thus, the model moment 0 maps of the C$^{18}$O (2--1) and SO (6$_{5}$--5$_{4}$) emission
are assumed to be single 2-dimensional Gaussians with the same axis ratio
($i.e.$, inclination angle $i$ = $\arccos \frac{8\arcsec}{16\arcsec}$ = 60$\degr$)
and the position angle.
The peak position of the model C$^{18}$O (2--1) moment 0 map
is fixed to be the protostellar position as measured from the 0.9-mm continuum
emission with the Natural weighting (Figure \ref{fig:contall}),
while that of the model SO (6$_{5}$--5$_{4}$)
moment 0 map the peak position of the observed SO moment 0 map derived from
the 2-dimensional Gaussian fitting.
The intrinsic emission extents
($i.e.,$ the size of the Gaussians) and the flux densities are unknown parameters,
as the observed images are processed through the interferometric filtering.
Thus, these two parameters for each of the C$^{18}$O and SO emission
are treated as free parameters. Hereafter we call these free parameters
``emission parameters''.

In contrast with the emission parameters,
parameters for gas motions should be common in the C$^{18}$O and SO models.
The model gas motions are divided into two regimes, the inner Keplerian-rotating
disk and the outer, rotating and infalling envelope.
Keplerian rotation motion with the central stellar mass of 0.1 $M_{\odot}$ is incorporated
within the Keplerian-disk radius, which is treated as a free parameter.
The spatial-velocity locations
of the $^{13}$CO (3--2) emission are consistent with those expected
from the Keplerian rotation with the central stellar mass of 0.1 $M_{\odot}$,
while the limited signal-to-noise ratio
and the spatial and spectral resolutions prevent us from investigating
the radial profile of the gas motion (see Figure \ref{fig:pvmaj}).
Outside the Keplerian disk a rotating and infalling protostellar envelope
is present. The rotating and infalling velocities of the envelope are simply assumed
to be uniform, and are regarded as free parameters.
In reality, the rotating and infalling velocities must have certain radial profiles
\cite{kra11,li14,tsu15b}. The spatial and spectral dynamic ranges of the present
SMA image cubes are, however, not high enough to discuss such radial profiles,
and in our model constant infalling and rotating velocities are adopted
as proxies of the averaged gas motions in the envelope.
Both forward- and counter-rotations of the disk with respect to the envelope rotation
are attempted in our model.
The internal velocity dispersion ($\equiv \sigma_{gas}$) of the envelope is
fixed to be 0.4 km s$^{-1}$ \cite{tak13,tak15},
whereas that of the disk is treated as a free parameter.
We call these free parameters to represent gas motions ``dynamical parameters''.

In our modeling,
the model moment 0 maps of the C$^{18}$O and SO emission are first produced
with a given set of the emission parameters. Then, with a given set of the dynamical
parameters, the model velocity channel maps of the C$^{18}$O and SO emission
are produced.
%
Since the model envelope and disk are assumed to be geometrically thin
and co-planar, with the given dynamical parameters the central velocity
and the line width at each position are determined.
It is thus straightforward to create model velocity channel maps of the C$^{18}$O
and SO emission.
The produced model velocity channel maps are converted into visibility data
with the same $uv$ coverage, field of view, and the spectral channels of the
real SMA observations. The model visibility data are then Fourier-transformed and CLEANed
with the same imaging parameters as those for the real observational data
(see Table \ref{lineobs}). In this interferometric modeling process, visibility noises
are not included for simplicity and for direct comparison with the observed images.
Finally the model interferometric velocity channel maps of the C$^{18}$O
and SO emission are compared to the real velocity channel maps, and the residual
velocity channel maps are made.
Two kinds of residual velocity channel maps are produced.
One is the observed velocity channel maps minus the model velocity
channel maps. This sort of the residual maps is straightforward for interpretation,
but is affected by the interferometric imaging process.
The other is made from the observed visibilities minus the model image components
or observed minus model visibilities. The subtracted visibilities are
Fourier-transformed without any deconvolution process.

Since there are a number of free parameters and our modeling includes SMA interferometric
sampling and imaging, it is not practical to perform full $\chi^2$ fitting of the model
velocity channel maps to the observed velocity channel maps. Instead, we manually changed
and adjusted the emission and dynamical parameters and tried to find a decent parameter set
which reasonably reproduces the observed velocity channel maps both in the C$^{18}$O and SO
emission. Strictly, a more complete search for the parameter space and proper
assessment of the $\chi^2$ value are required to prove which model
(forward- or counter-rotating models) is quantitatively better.
Thus, our modeling effort cannot definitely show which model is statistically better,
but just show a possible, plausible representation of the observed data.
Our searching results are summarized in Table \ref{models}.

\begin{deluxetable}{lll}
\tablecaption{Model Parameters\label{models}}
\tablewidth{250pt} 
\tablehead{\colhead{Parameter} &\colhead{Value} &\colhead{Fixed/Searched\tablenotemark{a}}}
\startdata
Systemic Velocity                              &$V_{LSR}$ = 6.8 km s$^{-1}$ &Fixed \\
Inclination $i$                       &60$\degr$                   &Fixed \\
Position Angle $\theta$               &154$\degr$                  &Fixed \\
Envelope $\sigma_{gas}$                        &0.4 km s$^{-1}$             &Fixed \\
Central Stellar Mass of the Keplerian Rotation &0.1 $M_{\odot}$             &Fixed \\
Keplerian Radius                               &200 au                      &Searched  \\
Disk $\sigma_{gas}$                            &1.0 km s$^{-1}$             &Searched \\
Envelope Infalling Velocity                    &0.16 km s$^{-1}$            &Searched \\
Envelope Rotational Velocity                   &0.2 km s$^{-1}$             &Searched   \\
FWHM size of the C$^{18}$O emission along the major axis &5$\farcs$2        &Searched \\
C$^{18}$O Integrated Flux Density              &3.9 Jy km s$^{-1}$          &Searched\\
FWHM size of the SO emission along the major axis        &2$\farcs$7        &Searched \\
SO Integrated Flux Density                     &2.8 Jy km s$^{-1}$          &Searched\\
\enddata
\tablenotetext{a}{``Fixed'' indicates the fixed model parameters. ``Searched'' indicates
the manually searched parameters from the comparisons between the model velocity channel
maps and the observed channel maps.}
\end{deluxetable}

\begin{figure}[ht!]
\figurenum{11}
\epsscale{1}
\begin{center}
\includegraphics[scale=0.55,angle=-90]{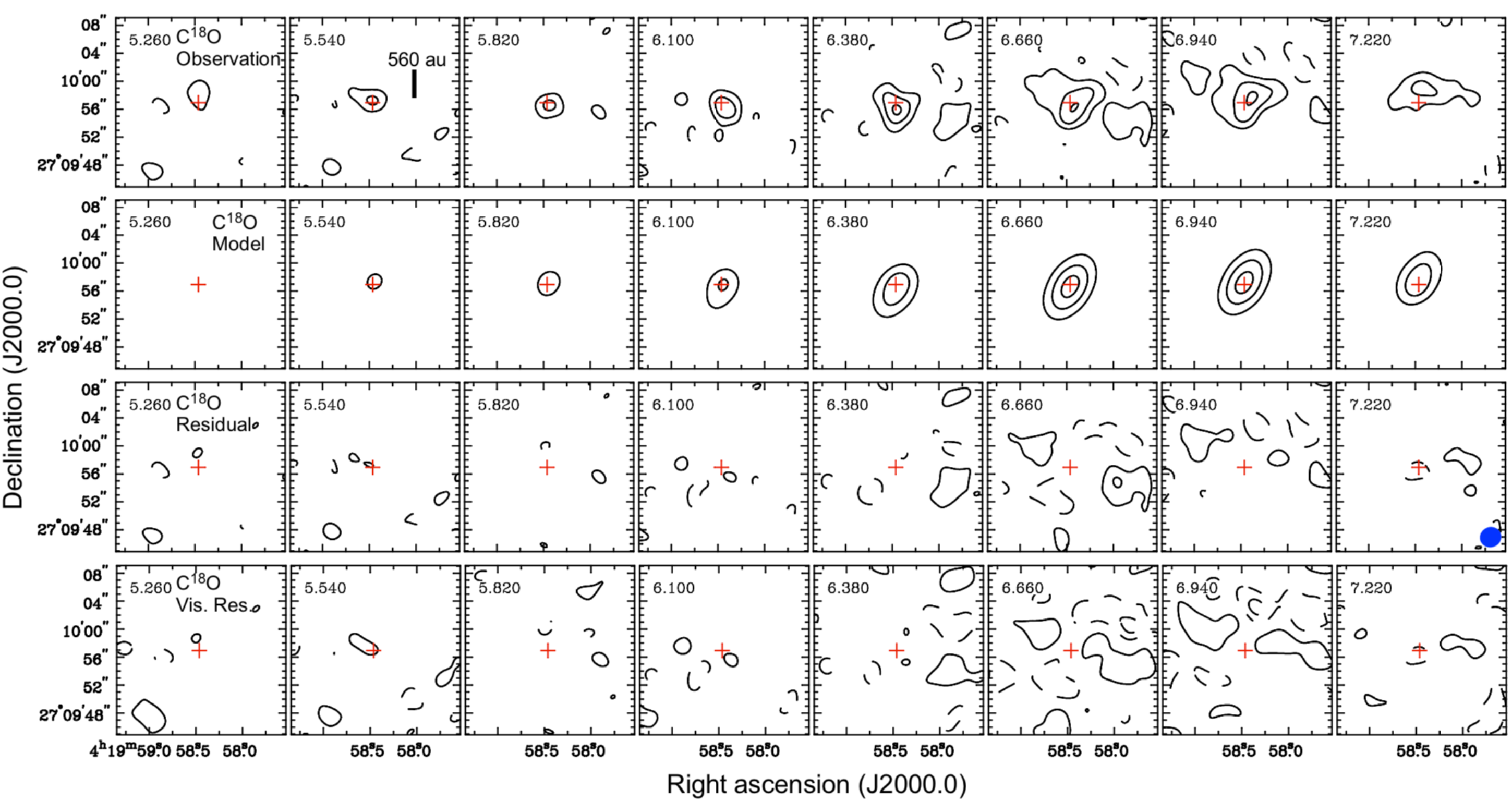}
\end{center}
\caption{Comparison of the observed and model velocity channel maps
of the C$^{18}$O (2--1) emission in I04169. Upper, upper-middle,
lower-middle, and lower panels
show the observed, model, residual velocity channel maps in
the image domain, and the velocity channel maps of the residual visibilities,
respectively. Contour levels are in steps of 2$\sigma$
(1$\sigma$ = 0.099 Jy beam$^{-1}$).
\label{fig:c18omod}}
\end{figure}

\begin{figure}[ht!]
\figurenum{12}
\epsscale{1}
\begin{center}
\includegraphics[scale=0.55,angle=-90]{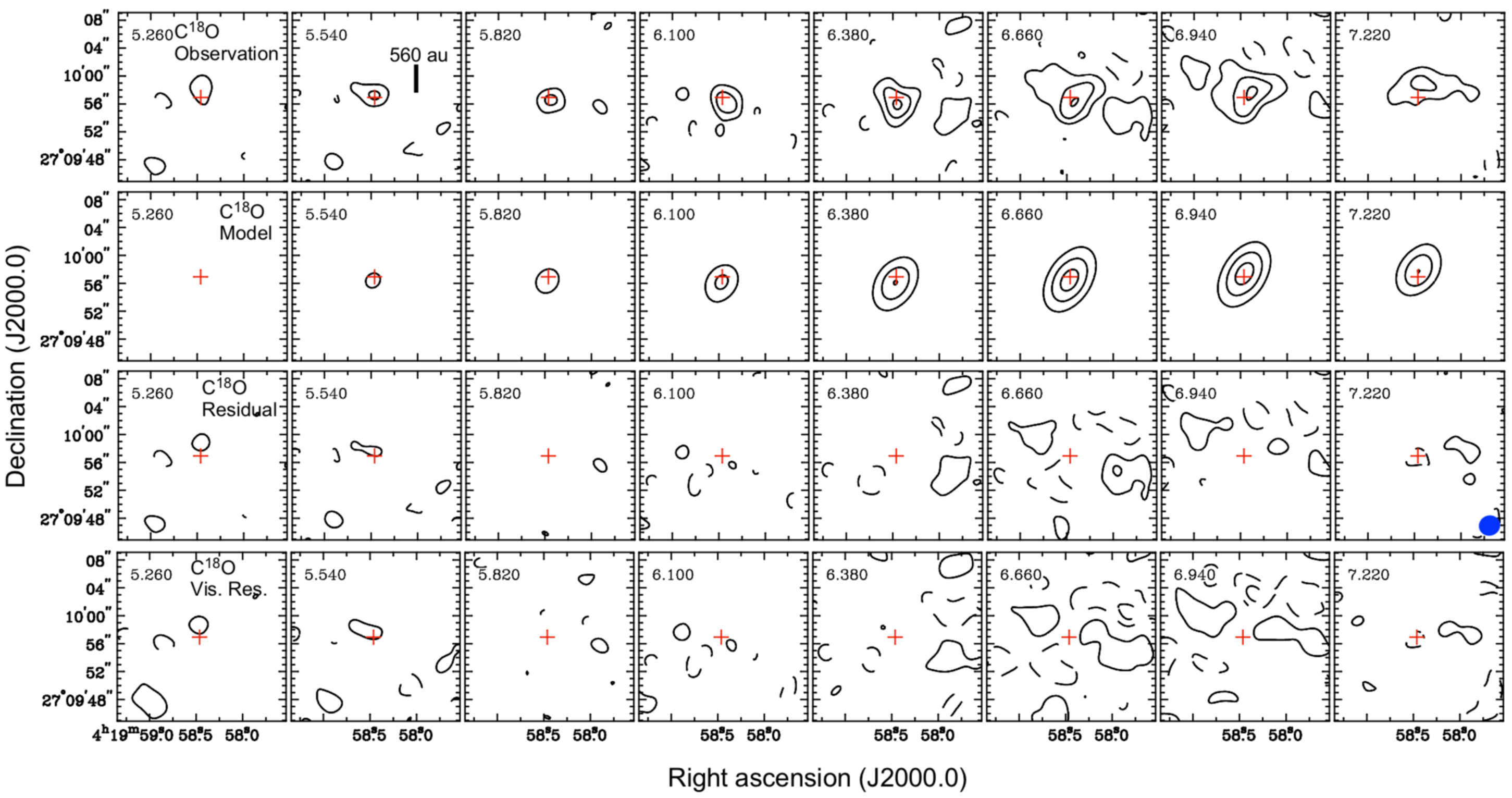}
\end{center}
\caption{Same as that of Figure \ref{fig:c18omod}, but for the model
of the forward rotation of the Keplerian disk.
\label{fig:c18omodjyun}}
\end{figure}

Figures \ref{fig:c18omod} and \ref{fig:c18omodjyun} compare the observed, model,
and the residual velocity channel maps of the C$^{18}$O emission for the counter-
and forward-rotating models, respectively. In both models the rotation and infall
of the envelope are common,
but the rotational direction of the central disk is opposite.
In the model velocity channel maps of
Figures \ref{fig:c18omod} and \ref{fig:c18omodjyun},
the disk components appear at $V_{LSR}$ = 5.54--5.82 km s$^{-1}$
to the north and south of the protostar, respectively.
Since the observed C$^{18}$O emission in these velocities are weak,
the flip of the rotational direction of the model disk does not produce
meaningful differences in the residual maps.
In the lower blueshifted velocities (6.10--6.66 km s$^{-1}$),
the emission centroid of both models appears to be located to the south of the protostar,
while in the redshifted velocities (6.94--7.22 km s$^{-1}$) to the north.
These features in the model C$^{18}$O velocity channel maps are
originated from the envelope, and the systematic emission shift
along the velocity traces that in the real observed velocity channel maps.
Since the bulk of the C$^{18}$O (2--1) emission is originated from
the envelope and the emission originated from the disk is minimal,
there is no significant difference between the counter- and forward-rotating models
in the residuals. 
The rms of the residual channel maps is 0.105 Jy beam$^{-1}$ and
0.125 Jy beam$^{-1}$ in the image-based and visibility-based residuals for
both models (The observational noise level is 0.099 Jy beam$^{-1}$.).

\begin{figure}[ht!]
\figurenum{13}
\epsscale{1}
\begin{center}
\includegraphics[scale=0.5,angle=-90]{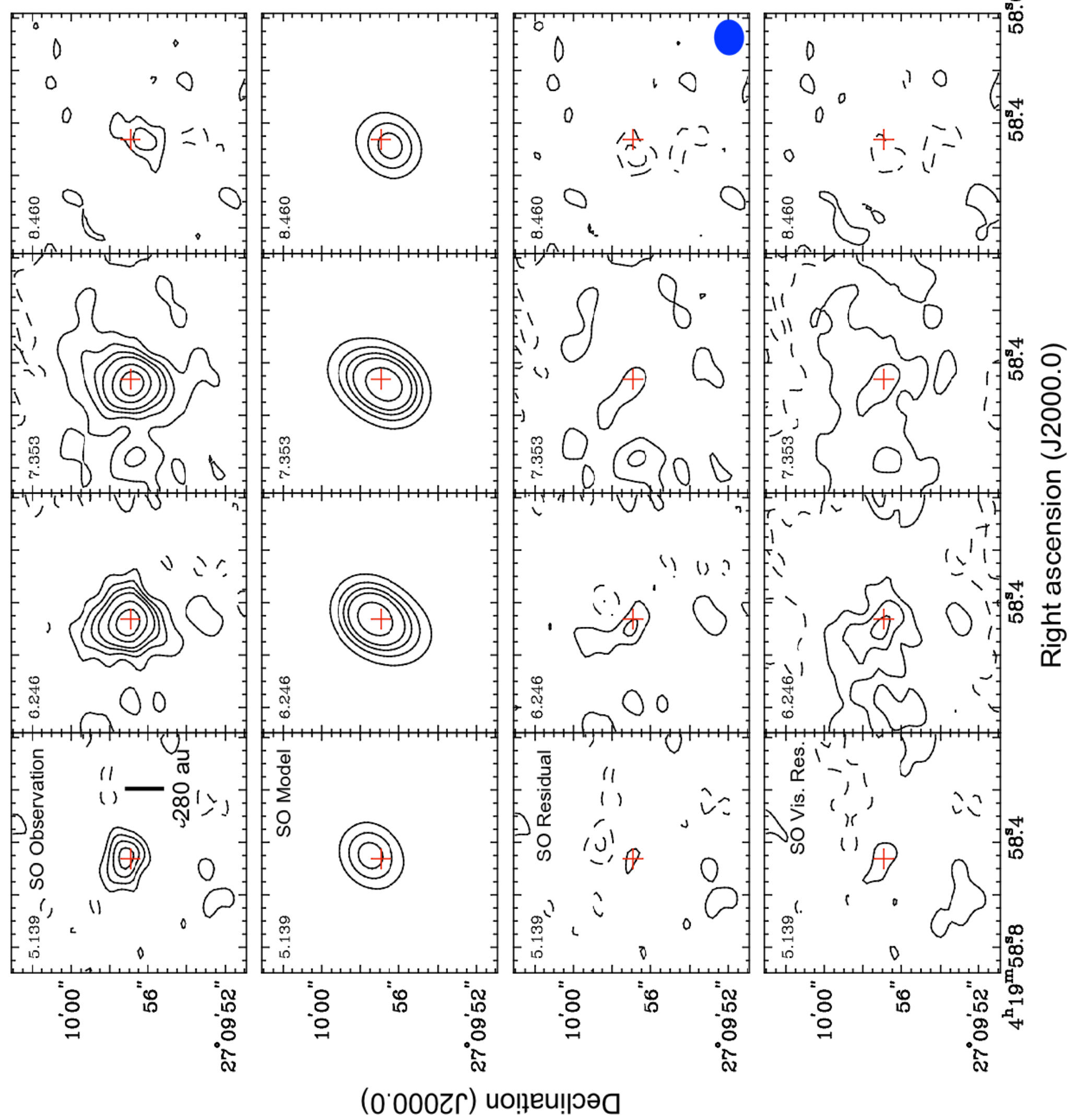}
\end{center}
\caption{Comparison of the observed and model velocity channel maps
of the SO (6$_{5}$--5$_{4}$) emission in I04169. Upper, upper-middle,
lower-middle, and lower panels
show the observed, model, residual velocity channel maps in
the image domain, and the velocity channel maps of the residual visibilities,
respectively. Contour levels are the same as those of Figure \ref{fig:soch}.
\label{fig:somod}}
\end{figure}

\begin{figure}[ht!]
\figurenum{14}
\epsscale{1}
\begin{center}
\includegraphics[scale=0.5,angle=-90]{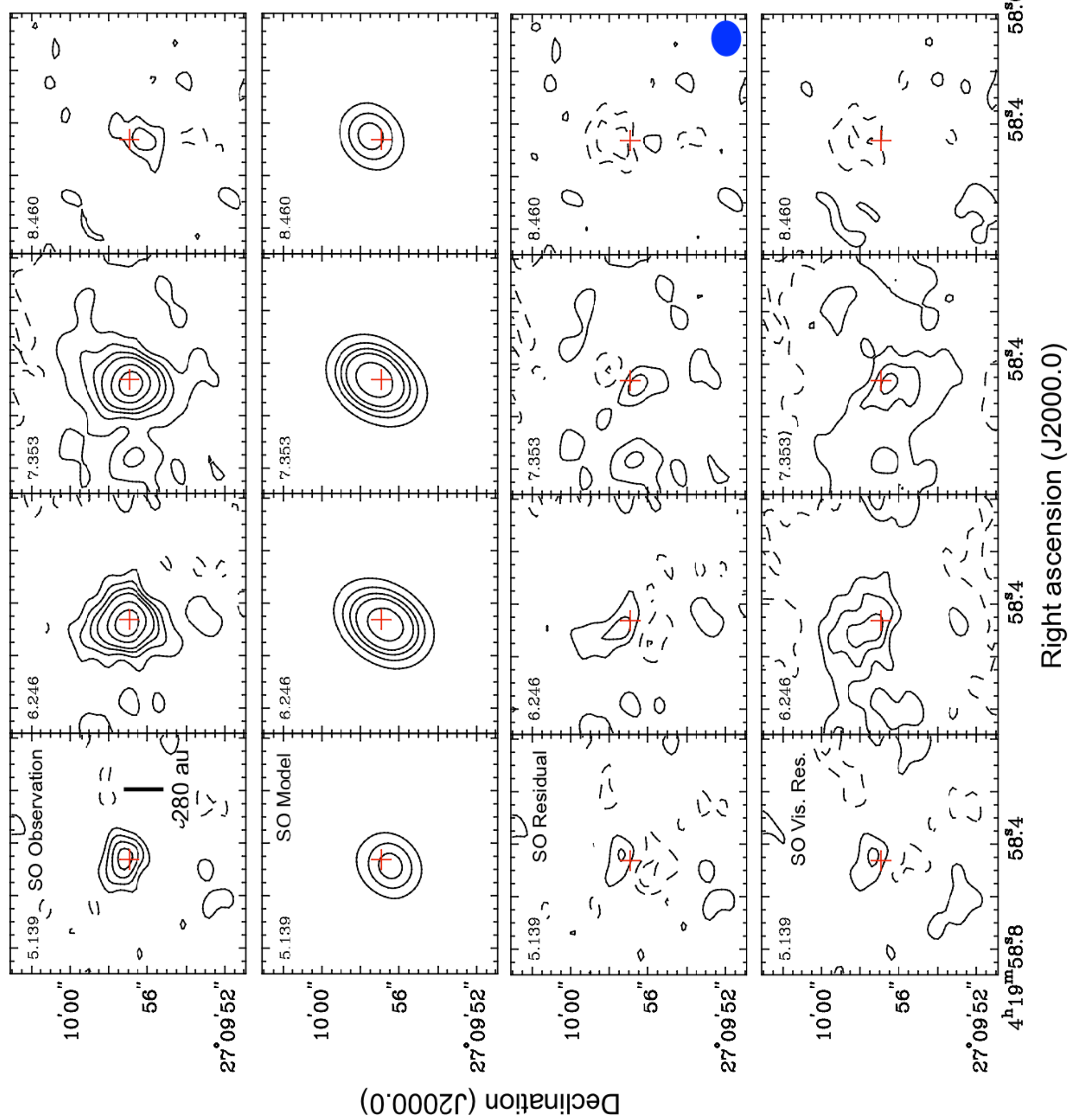}
\end{center}
\caption{Same as that of Figure \ref{fig:somod}, but for the model
of the forward rotation of the Keplerian disk.
\label{fig:somodjyun}}
\end{figure}

Figures \ref{fig:somod} and \ref{fig:somodjyun} show
the observed, model,
and the residual velocity channel maps of the SO emission for the counter-
and forward-rotating models, respectively. The observed SO velocity gradient
is opposite to that of the low-velocity C$^{18}$O emission.
The model with the disk rotational direction opposite to that of the envelope
(Figure \ref{fig:somod}) reproduces
the observed emission peak shift from north (blueshifted) to south (redshifted).
On the other hand, the forward-rotation model shows a completely opposite sign
of the velocity gradient (Figure \ref{fig:somodjyun}). The residual images show
``mirror-symmetric" residual patterns, and
in such a sense the counter-rotating model appears a better representation
of the SO velocity channel maps.
%
Note that the spatial resolution in the SO velocity channel maps
is higher than
that of the C$^{18}$O velocity channel maps shown
in Figures \ref{fig:c18omod} and \ref{fig:c18omodjyun}.
%
The velocity resolution of the SO velocity channel maps
is, on the other hand, much coarser than that of the C$^{18}$O velocity channel maps.
These observational differences reproduce the apparent differences of the image
cubes in the C$^{18}$O and SO emission.
In the image domain the
rms of the residual SO channel maps is 0.033 Jy beam$^{-1}$ for
the counter-rotating model
and 0.054 Jy beam$^{-1}$ for the forward-rotating model
(the observational noise level is 0.027 Jy beam$^{-1}$.).
In the visibility domain, the rms are
0.046 Jy beam$^{-1}$ and 0.048 Jy beam$^{-1}$ for
the counter-rotating and forward-rotating models, respectively.
The rms of the residual maps made from the visibility subtraction
does not show a significant difference between the counter-rotating
and forward-rotating models, as neither model is a perfect
representation of the observations derived from the complete
parameter search. In such a sense, our modeling effort cannot prove
that the counter-rotating model is statistically better than
the forward-rotating model.

To further investigate the model and observational results, the model P-V diagrams
along the major and minor axes are shown in Figures \ref{fig:pvmaj} and \ref{fig:pvmin},
respectively.
Comparisons between Figures \ref{fig:pvmaj}a
and c and between Figures \ref{fig:pvmaj}b and d show that the shapes of the observed P-V
diagrams resemble those of the counter-rotating model, which exhibit ``tilted diamond'' shapes.
The forward-rotation model shown in Figures \ref{fig:pvmaj}e and f does not reproduce
such features in the P-V diagrams, and the P-Vs of the forward-rotation model are always
skewed toward the two opposite quadrants. Along the minor axis, the observed
C$^{18}$O P-V diagram exhibits a slight velocity gradient. This velocity gradient
can be reproduced with a slow infalling velocity of 0.16 km s$^{-1}$ in the model. This infalling
velocity corresponds to the free-fall velocity with the central mass of 0.01 $M_{\odot}$
at $r =$700 au (see dashed curves in Figure \ref{fig:pvmin}). The observed and model
P-V diagrams of the SO emission along the minor axis do not show any clear velocity gradient,
consistent with our model interpretation that the SO emission primarily traces the central
disk component.

\begin{figure}[ht!]
\figurenum{15}
\epsscale{1}
\begin{center}
\includegraphics[scale=0.5,angle=-90]{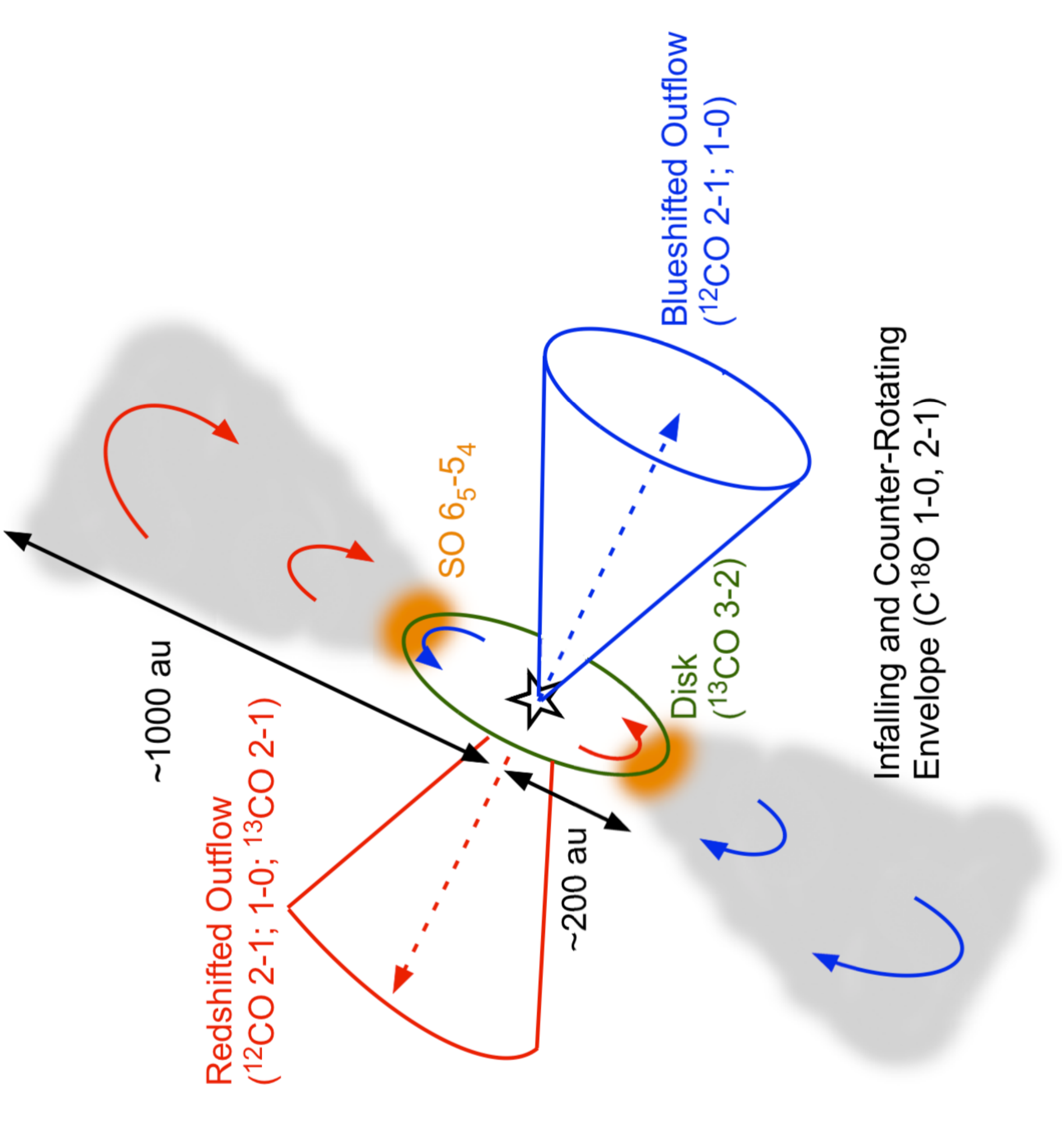}
\end{center}
\caption{Schematic picture of I04169 inferred from the present SMA observations.
\label{fig:scheme}}
\end{figure}

Our model results are not from thorough numerical parameter fitting
but obtained from manual parameter tuning. Thus our counter-rotating model
is not guaranteed to be an unique, best-fit solution.
There are no significant differences
of the residual rms between the forward- and counter-rotating models,
and in such a sense our modeling effort failed to show that
the counter-rotating model is statistically better than
the forward-rotating model.
Nevertheless, the counter-rotating model reproduces the observed velocity
gradient in the SO emission and the difference from that of the C$^{18}$O
emission, which is not reproduced with the forward-rotating model.
As the presence of the opposite velocity gradients is likely significant,
our counter-rotating model can still provide one possible, plausible
interpretation of the observed velocity structures.
Figure \ref{fig:scheme} shows a schematic picture of the
counter-rotating model.

\section{Discussion} \label{sec:dis}
\subsection{Nature of the Detected Velocity Features} \label{subsec:nature}

The molecular-line data toward I04169 reveal distinct velocity features
at different spatial scales. At $r \sim$400 -- 1000 au,
the northwestern part is redshifted and the southeastern part blueshifted,
as traced by the C$^{18}$O (2--1; 1--0) lines (see Table \ref{sumvel}).
In the inner $r \lesssim200$ au
the northwestern part becomes blueshifted and the southeastern part redshifted,
as seen in the SO (6$_5$--5$_4$) and $^{13}$CO (3--2) emission.
Our toy model shows that the observed velocity features
can be reproduced with a system of the central disk plus the outer counter-rotating,
infalling envelope, although our simple modeling
effort cannot prove that the counter-rotating
model is significantly better than the forward-rotating model in a quantitative way.

%

A possible source to change directions of velocity gradients is turbulence.
Recent theoretical simulations of magnetized and turbulent collapsing dense
cores show that turbulences produce additional angular momenta
and induce magnetic diffusions and reconnections \cite{san12,joo13,sei13,mat17}.
These effects reduce relative strengths of magnetic braking and thus promote
disk formation around the central stars.
The axis of the formed disk (or the angular-momentum vector)
can be misaligned from the that of the magnetic field, outflow, and the envelope
\cite{mat17}.
Observationally, Harsono et al. (2014) conducted PdBI observations of disks
around low-mass protostars,
and compared the rotational directions of the $r \sim$100 au scale disks
to the directions of the velocity gradients of the dense cores as seen in
the N$_{2}$H$^{+}$ emission observed with FCRAO \cite{cas02}.
The comparison between the $r \sim$100 au scale disks and $\sim$10000 au scale
cores shows that there are indeed differences of the directions of the velocity gradients
between the disks and cores, and the differences range from $\sim$90$\degr$ to
$\sim$230$\degr$.
Such differences of the velocity gradients between the large-scale cores and
disks can be attributed to the effect of the turbulences.
It is also possible that the identified change of the direction of the
velocity gradients in I04169 is due to the turbulence.
Presence of multiple gas clumps, which have different motions and velocities,
can also reproduce the observed, apparent flip of the velocity gradients in
I04169.

If the interpretation of the counter rotation is correct,
the effect of magnetic fields is a promising source to realize
such a velocity structure.
The physical mechanism to produce
such a counter rotation through magnetic fields is known as
the Hall effect, one of the non-ideal MHD effects \cite{war04,war12,bra12a,bra12b}.
In collapsing cloud cores, the Hall effect induces
toroidal (or azimuthal component of) magnetic fields from
poloidal magnetic fields at the mid-plane of the pseudo-disk or the flattened envelope.
The induced toroidal field exerts the magnetic
torque and the gas rotation with the left-handed screw direction
of the global poloidal field in the case of the negative Hall resistivity.
If the Hall induced magnetic torque is large enough
and has an opposite direction to the initial rotation
(this corresponds to the case in which the
poloidal magnetic field direction is parallel to the angular momentum
of the cloud core), the gas rotation can flip.
The counter rotating structure also appears even when the 
direction of the Hall induced magnetic torque is the same
as that of the initial rotation.
In this case, the midplane of the envelope
and the disk maintains the initial rotation direction,
but the upper envelope exhibits the counter rotation due to
the back reaction of the Hall induced forward rotation at the midplane.
In both cases, 
the flattened envelope or peudo-disk with the scale of $\sim 100$ AU
can exhibit the counter rotation \cite{kra11,li11,tsu15b,wur16,tsu17}.
Whether the Hall effect can induce the counter rotation or not
depends on the magnetic field strength of the parent cloud core. 
Previous theoretical studies have shown that the counter rotation is
realized in the cores with the mass-to-flux ratio
$\lambda \sim$5 \cite{li11,tsu15b,tsu17}. Here $\lambda$ is
normalized by its critical value. The Hall effect is more effective
under a stronger magnetic field ($i.e.$, lower $\lambda$),
and the previous observations of the magnetic fields in dense cores
have found $\lambda \sim$2 \cite{tro08,cru12}.
Thus, the flip of the rotational direction caused by the Hall effect
is likely possible under a typical condition of cloud cores.
The magnetic field strength can be estimated
from given $\lambda$ as $B=7.6 \times10^{-21} N(H_2) \lambda^{-1}~\mu G$
\cite{tro08}.
Toward I04169 $N(H_2)$ value is estimated to be 1.4$\times$10$^{22}$ cm$^{-2}$
\cite{mot01}, and $\lambda \sim$2 yields the field strength of 54 $\mu$G.

The limited spatial resolution and dynamic range of the present SMA data,
however, prevent us from discriminating these different interpretations.
If further higher-dynamic range observations of I04169 unveil
the consistency of the velocity gradient
from $r \sim$1000 au down to 400 au scale and then
the flip of the velocity gradient below 400 au,
such results must be a strong evidence for
the presence of counter rotation between the disk and envelope
and the magnetic-field origin.
In these spatial scales, theoretical simulations show
that the possible range of the misalignment between the central disk and the outer envelope
originated from turbulence is at most $\sim$30$\degr$ \cite{mat17}. Unless the core mass is
as high as 100 $M_{\odot}$, the misalignment
does not show a complete flip $i.e.$, 180$\degr$ \cite{sei13}.
Thus, if the future higher-dynamic range observations unveil
a systematic, consistent velocity structure and a flip
of the velocity gradient simultaneously, that can rule out
the origin of turbulence or multiple gas components.
In addition, a more thorough theoretical modeling (not toy model)
and statistical parameter search are required to prove that
the counter rotation caused by the magnetic effect is the most probable
and unique interpretation.

\subsection{Implications of the Opposite Velocity Gradients} \label{subsec:impli}

The present SMA observations of the Class I protostar I04169 have found
that at different spatial scales the directions of the velocity gradients are
opposite.
With our simple model, we suggest that one of the intriguing interpretations
is counter rotation between the protostellar envelope and disk caused by the magnetic fields,
whereas we admit that at this stage we cannot exclude the other possible explanations.

Furthermore, a marginal, slow infalling velocity ($\sim$0.16 km s$^{-1}$)
in the envelope has been identified. 
The identified infalling velocity corresponds to the free-fall velocity
toward the central mass of $\sim$0.01 $M_{\odot}$ or smaller at $r \lesssim$700 au.
If the mass of the central protostar is 0.1 $M_{\odot}$ as inferred from
the $^{13}$CO (3--2) and SO (6$_{5}$--5$_{4}$) results and our modeling,
the observed infalling velocity
is much smaller than the corresponding free-fall velocity.
Such small infalling velocities around the central disks have also been seen
in the other protostellar objects from our recent ALMA observations \cite{oha14,aso15}.
These results have been interpreted as the transitions from the infalling envelopes
to the central disks with the increasing centrifugal support. 
The physical origin of the slow
infalling velocity may also be the effect of magnetic fields
\cite{li11,ma11b}.


Opposite velocity gradients at different spatial scales
have also been seen in the other protostellar sources.
In HL Tau, the $r \sim$100 au scale protoplanetary disk exhibits blueshifted
emission to the southeast of the protostar and redshifted emission to northwest \cite{alm15}.
On the contrary, the follow-up ALMA observations by
Yen et al. (2017b) have found that the southeastern
part of the $r \sim$1000 au envelope around the protoplanetary disk of HL Tau
as seen in the $^{13}$CO (2--1) emission
is redshifted and the northwestern part blueshifted.
While Yen et al. (2017b) argued that a simple counter-rotating model is not sufficient
to fully reproduce the observed gas motions, presence of opposite
signs of the velocity gradients is identified with the ALMA observations.
Among the protostellar sample investigated by Harsono et al. (2014),
L1527 IRS shows almost a complete flip ($\sim$177$\degr$) of the
velocity gradient between the core and the disk.
Tobin et al. (2011) have also found that
the northern part of the $r \sim$8000-au scale protostellar envelope around
L1527 IRS is
blueshifted and the southern part redshifted, and that in the inner $r \sim$1000-au scale
the direction of the velocity gradient flips. The larger-scale velocity gradient
is consistent with the result from the single-dish C$_{3}$H$_{2}$
(2$_{12}$--1$_{01}$; 2$_{02}$--1$_{11}$) observations \cite{tak01},
and interferometric observations of L1527 IRS have confirmed the presence of the $r \lesssim 100$ au scale
Keplerian disk with the opposite velocity gradient \cite{tob12,sak14,oha14,aso17}.
In these two cases the differences of the spatial scales are within an order of magnitude.
These results imply that our SMA results of I04169 are not unique.

%

While both turbulences and magnetic fields have been considered to play a vital role in star and
circumstellar-disk formation out of cloud cores, 
it has been difficult to observationally identify such effects in protostellar sources.
Radial rotational profiles in
the protostellar envelopes
have been measured observationally to study gas motions into circumstellar-disk
formation ($e.g.$, Harsono et al. 2014; Yen et al. 2017a).
Whereas these observations show that the radial rotational profiles in the
envelopes and disks can be approximated to be $v_{rot} \sim r^{-1}$ ($i.e.$, rotation
with the conserved specific angular momenta) and $\sim r^{-0.5}$ (Keplerian rotation),
the rotational profiles measured from these observations are
not accurate enough to be directly compared with those from
theoretical simulations including magnetic fields and turbulence.
Thus, it is not straightforward to infer the impact of
magnetic fields and turbulence from the observed rotational profiles, 
if the direction of the rotational vector is common in
the envelopes and disks.
By contrast, the flip of the velocity gradient, if present,
is rather easy to identify observationally.
We thus suggest that further (re-)investigation
of the opposite velocity gradients of molecular gas around protostellar sources
should shed new light on the studies of star and circumstellar-disk formation.


\section{Summary} \label{sec:sum}

We have made high-resolution ($\sim$0$\farcs$5) SMA observations of the
Class I protostar I04169 in the $^{13}$CO (3--2) line and the 0.9-mm dust
continuum emission. We have also reduced and imaged the SMA archival data
of I04169 in the $^{12}$CO (2--1), $^{13}$CO (2--1), C$^{18}$O (2--1)
and the SO ($J_N$ = 6$_5$--5$_4$) lines and the 1.3-mm continuum emission
at angular resolutions of 2$\arcsec$-3$\arcsec$.
Compilation of these SMA data provides us with the following new insights
of the circumstellar materials around I04169.

\begin{itemize}
\item[1.] Both the 0.9-mm and 1.3-mm continuum emission are barely resolved,
and the extent of the 0.9-mm continuum emission is $\lesssim 0\farcs3 \sim$40 au.
The spectral index $\beta$ is estimated to be low ($\sim$0-0.5),
suggesting presence of dust growths.
The mass of the dusty component is estimated to be 0.0042 - 0.024 $M_{\odot}$
for $T_d$ = 10 - 30 K.
The redshifted $^{12}$CO (2--1) emission exhibits a tilted $U$-shaped feature with
its symmetric axis pointing toward the northeast of the protostar, and the blueshifted
$^{12}$CO (2--1) emission $V$-shaped feature toward the southwest.
The $^{12}$CO (2--1) spectrum toward the protostellar position exhibits
a blueshifted high-velocity wing until $V_{LSR} \lesssim$0.0 km s$^{-1}$.
 These results suggest that the $^{12}$CO (2--1) emission traces
the molecular outflow driven from I04169. The position angle of the outflow axis is
almost orthogonal to that of the $r \sim$1000-au protostellar envelope associated with I04169
as seen in the C$^{18}$O (1--0) emission (P.A.=154$\degr$).
The redshifted $^{13}$CO (2--1) emission appears to trace the same redshifted outflow
component as that traced by the $^{12}$CO (2--1) emission, while the blueshifted
$^{13}$CO (2--1) emission a distinct, compact feature to the northwest of the protostar.

\item[2.] The high-resolution $^{13}$CO (3--2) image cube shows that the blue-
($V_{LSR}$ = 3.7 - 5.9 km s$^{-1}$)
and redshifted (7.5 - 9.3 km s$^{-1}$) emission are located to the northwest and
southeast of the protostar, respectively,
with the outermost emission extent of $r \sim$100 au. The direction of the $^{13}$CO (3--2)
velocity gradient is almost orthogonal to that of the associated molecular outflow,
and thus along the major axis.
The peak locations of the blueshifted and redshifted SO (6$_{5}$--5$_{4}$) emission
are consistent with those of the $^{13}$CO (3--2) emission.
On the other hand, the lower-resolution C$^{18}$O (2--1) image cube exhibits distinct
velocity features in the low-velocity range (5.7 - 7.6 km s$^{-1}$):
The blue- and redshifted C$^{18}$O (2--1) emission are located to the south
and north of the protostar, respectively, with the outermost extent of $r \sim$400 au.
Along the major axis (NW-SE), the sign of the velocity gradient of the C$^{18}$O (2--1)
emission is opposite to that of the $^{13}$CO (3--2) and SO (6$_{5}$--5$_{4}$) emission,
but is consistent with that of the $r \sim$1000-au scale protostellar envelope.
Such different velocity structures in the different molecular lines
are also confirmed from the direct fitting to the visibility data in
the individual velocity channels.
Along the minor axis the C$^{18}$O (2--1) emission shows a marginal
NE (red) - SW (blue) velocity gradient too.
In the highly blueshifted velocity (5.3 - 5.5 km s$^{-1}$),
the C$^{18}$O (2--1) counterpart of the
blueshifted $^{13}$CO (3--2) emission is also present.

\item[3.] To interpret the observed velocity structures,
we have constructed toy models of an infinitesimally-thin Keplerian disk
plus envelope in the C$^{18}$O (2--1) and SO (6$_{5}$--5$_{4}$) emission, including
the SMA imaging simulations. We tested models where the envelope and the inner
disk are co-rotating and counter-rotating with each other.
Our manual parameter search has found that the counter-rotating model better
reproduces the observed velocity gradient in the SO emission.
On the other hand, since our manual parameter search failed to obtain
statistically significant fitting results with neither the counter-rotating
and forward-rotating models, we could not verify that the counter-rotating model
is significantly better than the forward rotating model.
As the presence of the opposite velocity gradients between
the C$^{18}$O and SO emission is likely significant,
our counter-rotating model can still provide one possible, plausible
interpretation of the observed velocity structures.
Higher sensitivity, and higher spatial dynamic-range
observations of I04169 along with a more detailed model, and a more complete exploration
of the parameter space associated with these models, are needed to definitively
demonstrate that a counter-rotating envelope/disk is a more proper
model than a standard co-rotating envelope/disk model.


\item[4.] The observed velocity structures in the circumstellar material
of I04169 can be interpreted as either turbulent gas motions, multiple gas clumps
at different velocities, or counter rotation between the disk and envelope.
With the present data we cannot define which is the most appropriate model.
Nevertheless, we suggest that counter rotation between the disk and envelope
caused by the effect of the magnetic fields
is one of the intriguing interpretations for the observed flip of the direction of the velocity gradient.
There are indeed other protostellar sources which exhibit the different velocity gradients
between the disks and envelopes.
While the effects of the magnetic fields have been considered to be critical
in star and circumstellar-disk formation, it has been difficult
to observationally identify such effects in protostellar sources.
The counter rotation could be a promising observational signature to investigate
the magnetic effects. Higher-sensitivity and higher spatial dynamic-range
observations of protostellar envelopes and disks, along with
(re-)investigation of the existing ALMA data of protostellar sources,
should provide us with important insights of the effects of the magnetic fields,
which have been difficult to identify observationally.
\end{itemize}


\acknowledgments
We would like to thank N. Ohashi and J. Lim for their
fruitful discussions, and all the SMA staff supporting this work.
S.T. acknowledges a grant from the Ministry of Science and Technology (MOST) of Taiwan
(MOST 102-2119-M-001-012-MY3), and JSPS KAKENHI Grant Numbers JP16H07086 and
JP18K03703 in support of this work.



\vspace{5mm}
\facilities{SMA}

\software{MIR, Miriad}

\listofchanges

\end{document}